\documentclass[onecolumn,useAMS,usenatbib]{mn2e}
\usepackage{graphicx}
\usepackage{natbib}

\usepackage{bm}
\usepackage{amsfonts}
\usepackage{latexsym}
\usepackage[latin1]{inputenc}
\usepackage{amsmath}
\usepackage{epsfig}
\usepackage{amsbsy}

\newcommand{\mtb}[1]{\mathbf{#1}}

\newcommand{\ord}{\mathrm o}
\newcommand{\s}{\mathrm s}
\newcommand{\n}{\mathrm n}
\newcommand{\p}{\mathrm p}

\newcommand{\x}{\mathrm x}
\newcommand{\y}{\mathrm y}
\newcommand{\crt}{\mathrm c}

\newcommand{\veps}{\varepsilon}

\def \nn  {\nonumber}

\def \eps{\epsilon}
\def \veps{\varepsilon}
\def\jnl@style{\it}
\def\aaref@jnl#1{{\jnl@style#1}}

\def\aaref@jnl#1{{\jnl@style#1}}

\def\aj{\aaref@jnl{AJ}}                   
\def\apj{\aaref@jnl{ApJ}}                 
\def\apjl{\aaref@jnl{ApJ}}                
\def\apjs{\aaref@jnl{ApJS}}               
\def\apss{\aaref@jnl{Ap\&SS}}             
\def\aap{\aaref@jnl{A\&A}}                
\def\aapr{\aaref@jnl{A\&A~Rev.}}          
\def\aaps{\aaref@jnl{A\&AS}}              
\def\mnras{\aaref@jnl{MNRAS}}             
\def\prd{\aaref@jnl{Phys.~Rev.~D}}        
\def\prl{\aaref@jnl{Phys.~Rev.~Lett.}}    
\def\qjras{\aaref@jnl{QJRAS}}             
\def\skytel{\aaref@jnl{S\&T}}             
\def\ssr{\aaref@jnl{Space~Sci.~Rev.}}     
\def\zap{\aaref@jnl{ZAp}}                 
\def\nat{\aaref@jnl{Nature}}              
\def\aplett{\aaref@jnl{Astrophys.~Lett.}} 
\def\apspr{\aaref@jnl{Astrophys.~Space~Phys.~Res.}} 
\def\physrep{\aaref@jnl{Phys.~Rep.}}      
\def\physscr{\aaref@jnl{Phys.~Scr}}       

\title[Hydrodynamics of superfluid neutron stars]{Hydrodynamics of
  rapidly rotating superfluid neutron stars with mutual friction}

\author[A. Passamonti $\&$ N. Andersson]
{A. Passamonti\thanks{E-mail:a.passamonti@soton.ac.uk} , N. Andersson
\\ \\
School of Mathematics, University of Southampton, Southampton SO17 1BJ, UK}

\begin{document}

\date{\today}

\pagerange{\pageref{firstpage}--\pageref{lastpage}} \pubyear{}

\maketitle

\label{firstpage}


\begin{abstract}

We study time evolutions of superfluid neutron stars, focussing on the
nature of the oscillation spectrum, the effect of mutual friction
force on the oscillations and the hydrodynamical spin-up phase of
pulsar glitches. We linearise the dynamical equations of a Newtonian
two-fluid model for rapidly rotating backgrounds. In the axisymmetric
equilibrium configurations, the two fluid components corotate and are
in $\beta$-equilibrium. We use analytical equations of state that
generate stratified and non-stratified stellar models, which enable us
to study the coupling between the dynamical degrees of freedom of the
system. By means of time evolutions of the linearised dynamical
equations, we determine the spectrum of axisymmetric and
non-axisymmetric oscillation modes, accounting for the contribution of
the gravitational potential perturbations, i.e. without adopting the
Cowling approximation. We study the mutual friction damping of the
superfluid oscillations and consider the effects of the
non-dissipative part of the mutual friction force on the mode
frequencies. We also provide technical details and relevant tests for
the hydrodynamical model of pulsar glitches discussed by
\cite*{2010MNRAS.tmp..554S}. In particular, we describe the method
used to generate the initial data that mimic the pre-glitch state, and
derive the equations that are used to extract the gravitational-wave
signal.

\end{abstract}


\section{Introduction} \label{sec:Int}

Mature neutron stars are expected to have superfluid and
superconducting components in their interior. Shortly after a neutron
star's birth the temperature decreases below $T\simeq10^{9}~\rm{K}$,
at which point superfluid neutrons should be present both in the inner
crust and the outer core, while the core protons should form a
superconductor. At all relevant temperatures, the electrons form a
``normal'' fluid that is tightly locked to the protons due to the
electromagnetic interaction. This suggests that the dynamics of mature
neutron stars depends on the detailed interaction between coupled
superfluids-superconductors~\citep*{2010arXiv1001.4046G}, i.e.
represents a complex physics problem. The situation is not expected to
simplify if one also accounts for the inner neutron star core, at
several times the nuclear saturation density, where exotic states like
hyperon superfluid mixtures or deconfined quark condensates may be
present.

Although it is generally appreciated that neutron stars have this very
complicated structure, the evidence for the presence of the different
superfluid phases remain indirect. The strongest support comes from
observed pulsar glitches, rapid spin-up events seen in a number of
young pulsars (and also some magnetars) during their magnetic
slow-down phase. The typical glitch size is very small, representing a
relative change ($\Delta \Omega$) in the observed rotation rate
($\Omega$) in the range $10^{-9} < \Delta \Omega / \Omega < 10^{-5}$.
The currently accepted model for these events relies on the transfer
of angular momentum between a (faster spinning) superfluid neutron
component and the star's (slower spinning) elastic crust (to which the
magnetic field is anchored).  The exchange is thought to be mediated
by neutron vortices (by means of which the superfluid mimics bulk
rotation) and the associated mutual friction~\citep{ALS84}.

A challenge for future observations is to probe the detailed physics
of a neutron star's interior. In this context, asteroseismology
associated with either gravitational or electromagnetic signals seems
particularly promising. In fact, the quasiperiodic oscillations seen
in the tails of giant magnetar flares may have provided us with the
first opportunity to test our theoretical models against observational
data~\citep[see for instance][and references
therein]{2007Ap&SS.308..625W}. The observed variability likely
originates from crustal oscillations and depends on the detailed crust
dynamics and the interaction with the neutron star's magnetic field.
These observations have led to a resurgence of interest in
neutron-star seismology and a renewed assault on the problem of
magnetic star oscillations, a seriously challenging problem from the
theory point-of-view~\citep*[see][for a discussion of the
literature]{2009MNRAS.396.1441C}.  In the context of the present
paper, the potential relevance of the neutron superfluid that
penetrates the neutron star crust is particularly
relevant~\citep*{2009MNRAS.396..894A,2009CQGra..26o5016S}.  The
prospect of detecting gravitational waves from oscillating neutron
stars is also exciting, especially since the associated signals will
allow us to probe the high-density region and hence the supranuclear
equation of state
(EoS)~\citep*{1998MNRAS.299.1059A,benhar-2004-70,2007MNRAS.374..256S,
2009arXiv0912.0384A}.

In order to faciliate future observations and the decoding of
collected data, we need to improve our models considerably. The
superfluid aspects are particularly interesting in this respect, since
the oscillation spectrum of a superfluid star is more complex than
that of a single fluid model. In superfluid regions fluid elements can
execute both co- and counter-moving motion, leading to the existence of
unique ``superfluid'' oscillation modes. Our understanding of the
nature of the additional degree(s) of freedom and the effect on
observables must be improved by detailed modelling, ultimately in the
context of general relativistic multi-fluid dynamics.

The present work presents recent progress towards this goal.  We study
the oscillations of superfluid neutron stars by evolving in time the
linearized two-fluid equations in Newtonian gravity. We improve on the
analysis of~\cite{2009MNRAS.396..951P} by including the perturbations
of the gravitational potential. We also account for the mutual
friction force associated with vortices, and implement quadrupole
extraction of the gravitational-wave signal associated with the fluid
motion.  We provide the detailed analysis (and relevant code tests)
for the configurations that we recently used to study the
hydrodynamics of pulsar glitches~\citep{2010MNRAS.tmp..554S}.  We
consider two simple analytical EoS and construct two distinct
sequences of rapidly rotating stars, the main difference being the
presence or absence of composition gradients.  Such gradients impact
on the superfluid dynamics, as the co- and counter-moving degrees of
freedom are coupled in stratified models.  From time-evolutions of the
relevant perturbation equations, with the gravitational potential
perturbation included, we determine the axi- and non-axisymmetric
oscillation modes for models that rotate up to the mass shedding
limit. Finally, we account for the (standard form of the) mutual
friction force. This adds two coupling terms to the equations of
motion.  One component is dissipative and damps an oscillation mode,
while the other modifies the frequencies of the superfluid modes. We
study both these effects and infer an analytical relation for the
associated frequency change of the non-axisymmetric superfluid
fundamental and inertial modes.

\section{Equations of Motion} \label{sec:pert-eqs}

In a basic model for superfluid neutron stars, the matter constituents
are superfluid neutrons, superconducting protons and normal
electrons. Given the typical dynamical timescale of stellar
oscillations, one would expect the charged particles to be efficiently
locked together by the electromagnetic interaction. Therefore, the
dynamics of superfluid stars depends on two components, a neutron
superfluid and a neutral conglomerate of protons and electrons. For
simplicity, we will refer to the latter mixture as the ``protons'' in
the following. More detailed discussion and justification for the
two-fluid model is provided by~\citet{1991ApJ...380..515M,
  1991ApJ...380..530M}, \citet{2004PhRvD..69d3001P}
and~\citet{2006CQGra..23.5505A}.

When the mass of each fluid component is conserved, i.e. when we
neglect the various particle reactions, the dynamics of a superfluid
star is described by two mass conservation laws, two Euler-type
equations and the Poisson equation for the gravitational
potential~\citep{2004PhRvD..69d3001P}. These take the form;
\begin{equation}
\partial_t \rho_{\x} + \nabla_{i} \left( \rho_{\x} v_{\x}^{i} \right)
= 0 \, , \label{eq:Mcon}\\
\end{equation}
\begin{equation}
\left( \partial_t + v_{\x}^{k} \nabla_{k} \right)
 \left( v_{i}^{\x} + \varepsilon_{\x} w_{i}^{\y\x} \right) + \nabla_{i} \left( \Phi + \tilde{\mu}_{\x} \right)
+ \varepsilon_{\x} w_{k}^{\y\x} \nabla_{i} v_{\x}^{k} =  \frac{f^{\x}_{i}}{\rho_{\x}} \, ,  \label{eq:Euler} \\
\end{equation}
\begin{equation}
 \nabla^2 \Phi = 4 \pi G \rho \, . \label{eq:Poisson}
\end{equation}
These equations are given in a coordinate basis, which means that the
indices $i$ and $k$ denote spatial components of the various
vectors. Meanwhile the indices $\x$ and $\y$ label the two fluid
components. In the present case these constituent indices will be $\n$
for the neutrons and $\p$ for the protons. Throughout this work, the
summation rule for repeated indices applies only for spatial
indices. In equations~(\ref{eq:Mcon})--(\ref{eq:Poisson}), the total
mass density is $\rho = \rho_\n + \rho_\p$, $\tilde \mu_{\x}$ is the
chemical potential for each fluid component (scaled with the particle mass
$m=m_\n=m_\p$), $\Phi$ is the gravitational potential, while the relative
velocity between the two fluids is $w^{\x\y}_{i} \equiv v_{i}^{\x} -
v_{i}^{\y}$.  The parameter~$\veps_{\x}$ accounts for the
non-dissipative entrainment effect. In a neutron star core the entrainment
is due to the strong interaction between the nucleons. From
equation~(\ref{eq:Euler}), it is clear that it leads to a
momentum that is not longer aligned with the individual component
velocity.  The vector field $\mtb{f}^{\x}$ represents the force
density acting on the $\x$ fluid component. In this paper we consider
only the vortex mediated mutual friction force.  The general form of
this force is
\begin{equation}
f^{\x}_{i} = 2 \rho_\n \left( \mathcal{B}' \, \eps_{ijk} \,
 \Omega^{j} w_{\x\y}^{k} + \mathcal{B} \, \eps_{ijk} \, \eps^{k}_{~lm}
 \, \hat{\Omega}^{j} \, \Omega^l \, w_{\x\y}^{m} \right) \, ,
\end{equation}
where $\hat{\Omega}^{i} = \Omega^{i} / \Omega$ represents the bulk
rotation (later we will assume that the two fluids co-rotate in the
unperturbed background), and $\mathcal{B}$ and $\mathcal{B}'$ are the
mutual friction parameters.

\subsection{Equation of State} \label{sec:EoS}

The equation of state (EoS), that is needed to close the system of
equations, can be described by an energy functional
\begin{equation}
\mathcal{E} = \mathcal{E} \left( \rho_\n, \rho_\p , w_{\n \p}^2
\right) \, , \label{eq:EoS}
\end{equation}
that ensures Galilean invariance.  The chemical potential
$\tilde{\mu}_\x$ and the entrainment parameter $\varepsilon_{\x}$ are
then defined by
\begin{eqnarray}
\tilde{\mu}_{\x} & \equiv & \left. \frac{\partial \mathcal{E}}{\partial \rho_{\x} }
\right| _{\rho_{\y}, w_{\x\y}^2}\, , \label{eq:defmu} \\
 \varepsilon_{\x} & \equiv & 2 \rho _{\x}\left.  \frac{\partial
\mathcal{E}}{\partial w^{2}_{\n\p} } \right|_{\rho_\x,\rho_\y} \, . \label{eq:vareps}
\end{eqnarray}
When the relative velocity between the two fluids is small, as is the case in most systems
of practical relevance,
equation~(\ref{eq:EoS}) can be expanded in a series:
\begin{equation}
\mathcal{E} = \mathcal{E}_{0} \left(\rho_\n, \rho_\p \right)
+ \alpha_0 \left( \rho_\n, \rho_\p \right) w_{\n \p}^2 + \mathcal{O}\left(w_{\n \p}^4\right) \, , \label{eq:EoSbulk}
\end{equation}
This has the advantage that the bulk EoS $\mathcal{E}_{0}$ and the entrainment parameter
$\alpha_0$ can
be independently specified at $\mtb{w}_{\n\p}=\mtb{0}$.  From
equation~(\ref{eq:vareps}) it follows that the entrainment parameter
$\varepsilon _{\x}$ is related to the function $\alpha_0$ by
\begin{equation}
\rho_\x  \varepsilon_\x = 2 \alpha_0 \, .  \label{eq:alp}
\end{equation}

Despite recent developments~\citep{2008MNRAS.388..737C}, we do not yet
have a realistic EoS that consistently describes the superfluid
properties of a neutron star core. Therefore, we consider two
analytical EoS, based on generalisation of the familiar $n=1$
polytrope. These models are particularly useful if we want to explore
the role of entrainment, composition stratification and symmetry
energy. Moreover, since the two EoS have been used elsewhere we have
``independent'' tests of our numerical results. The main difference
between our two sets of models is the presence, or absence, of composition
gradients. This is important since the co- and counter-moving degrees
of freedom are coupled in stratified neutron stars, which means that
the gravitational-wave spectrum may contain the imprints of
``superfluid'' modes (see Sec.~\ref{sec:Res}). This would not be the
case in a non-stratified model.

The first EoS is determined by the following
expression~\citep{2002A&A...381..178P,2004MNRAS.347..575Y,
  2009MNRAS.396..951P}:
\begin{equation}
\mathcal{E}_{0} =   \frac{K}{1-\left( 1 + \sigma \right) x_\p} \rho_\n ^2
- \frac{2 K \sigma }{1-\left( 1 + \sigma \right) x_\p} \rho_\n \rho_\p
+ \frac{ K  \left[ 1 + \sigma - \left( 1 + 2 \sigma \right) x_\p \right] }
{x_\p \left[  1-\left( 1 + \sigma \right) x_\p \right] } \rho_\p^2 \label{eq:EoSA} \, ,
\end{equation}
where $K$ is a polytropic constant, $x_{\p}$ is the proton fraction
and $\sigma$ is a parameter that can be related to the symmetry
energy~\citep{2002A&A...381..178P}. In this EoS, both $x_{\p}$ and
$\sigma$ are taken to be constant~\citep{2009MNRAS.396..951P}.  Using
equation~(\ref{eq:EoSA}), we construct a sequence of co-rotating
axisymmetric configurations without composition gradients. These
correspond to the A models used by~\citet{2009MNRAS.396..951P}.

In order to study the effects of stratification on the oscillation
spectrum we consider a second EoS, defined by
~\citep{2002A&A...393..949P, 2002PhRvD..66j4002A,
  2009MNRAS.396..951P}:
\begin{equation}
\mathcal{E}_{0} = k_{\n} \, \rho_\n^{\gamma_n} + k_{\p} \,
\rho_\p^{\gamma_\p} \, . \label{eq:EosPR}
\end{equation}
Here, the coefficients $k_\x$ and $\gamma_\x$ are constants. We
consider $\gamma_{\n} = 1.9$, $\gamma_{\p}=1.7$ for all the rotating models.
In the numerical code, the coefficients $k_\x$ are given in
units of $G R^{2}_{eq} \rho_{o}^{2-\gamma_\x}$, where $G$ is the
gravitational constant and $R_{eq}$ is the equatorial radius of the
stellar model. We take them to have the values $k_\n =0.682$ and $k_\p =3.419$
for  the non-rotating model, which corresponds to model III used
by~\citet{2002A&A...393..949P}. Note that for rotating models, the dimensionless $k_\x$
can assume a different value with respect to the non-rotating star.
For instance, when we impose that the central proton fraction is constant  for all the sequence of rotating models
(see Section~\ref{sec1}).
From equations~(\ref{eq:defmu})
and~(\ref{eq:EosPR}) it follows that the chemical potential and mass
density  are related by
\begin{equation}
\rho_\x = \left( \frac{\tilde \mu _\x }{k_\x \gamma_\x } \right)
^{N_\x} \, ,   \label{eq:rhomu}
\end{equation}
where the polytropic index is given by $N_\x = \left( \gamma_{\x} -1
\right) ^{-1}$. From this result we can determine the proton fraction for a given stellar
 model by imposing  $\beta$-equilibrium. After some calculations, we obtain:
\begin{equation}
x_\p = \left[ 1 + \frac{ \left( \gamma_\p k_\p \right) ^{N_\p}
}{\left( \gamma_\n k_\n \right) ^{N_\n}} \, \tilde \mu ^{N_\n-N_\p}
\right]^{-1} \, .  \label{eq:xp}
\end{equation}
This EoS will be used to construct a sequence of stratified rotating
models, where the central proton fraction is fixed, $x_\p (0)=
0.1$. These equilibrium configurations have already been used
by~\cite{2010MNRAS.tmp..554S}, and will be refered to as models C in
the following.

\section{Equilibrium configurations} \label{sec:eq-conf}

We study the oscillations of rotating axisymmetric background models where
neutrons and protons are in $\beta$-equilibirum and co-rotate with
constant angular velocity, i.e. we have $\Omega_\n=\Omega_\p$. In this work, we also
assume that two fluid components coexist
throughout the stellar volume. This is obviously artifical; the outer region of a real neutron star
will not be superfluid. However, at this stage our main interest is in the bulk core dynamics.
In the future we plan to extend our model to account appropriately for the
expected superfluid regions. At that point we will also consider the role of the elastic crust.

In Sec.~\ref{sec1}, we introduce the  equations that govern stationary
co-rotating equilibrium configurations and solve them numerically for
the EoS~(\ref{eq:EoSA}) and (\ref{eq:EosPR}).  In
Sec.~\ref{sec:Noncorback}, we then describe the perturbative approach
developed by~\cite{2004MNRAS.347..575Y} for determining stationary
configurations in which the two fluids rotate with a small velocity
lag. With this method we can obtain non-corotating models as small deviations from a co-rotating equilibrium.
Subsequently, we use this approach to determine the
initial conditions for hydrodynamical glitch evolutions.

\begin{table}
\begin{center}
\caption{\label{tab:back-models} This table provides the main
  parameters for the two sequences of rotating models.  The first column labels each model.
  In the second and third columns
  we give, respectively, the ratio of polar to equatorial axes and the
  angular velocity of the star. In the fourth column, the rotation rate
  is compared to the Kepler velocity $\Omega_K$ that represents the
  mass shedding limit. The ratio between the rotational kinetic energy
  and gravitational potential energy $T/|W|$ and the stellar mass are
  given in the fifth and sixth columns, respectively. Finally, the seventh column gives the value of the chemical potential at the centre of the star. All quantities
  are given in dimensionless units, where $G$ is the gravitational
  constant, $\rho_0$ represents the central mass density and $R_{eq}$
  is the equatorial radius.}
\begin{tabular}{ c  c  c c c c c  }
\hline
 Model &   $ R_p / R_{eq} $  &  $ \Omega / \sqrt{G\rho_0}$ & $\Omega / \Omega_K$  & $ T/|W| \times 10^{2}$
  & $ M / (\rho_0 R_{eq}^3)$ & $ \tilde \mu_0 / (G \rho_0 R_{eq}^2)$  \\
\hline
 A0 &  1.00000   &   0.00000   &  0.00000  &  0.00000  &  1.2732   & 1.2732 \\
 A1 &  0.99792   &   0.05913   &  0.08153  &  0.05802  &  1.2701   & 1.2701 \\
 A2 &  0.98333   &   0.16675   &  0.22992  &  0.38482  &  1.2479   & 1.2477 \\
 A3 &  0.95000   &   0.28729   &  0.39613  &  1.16918  &  1.1967   & 1.1962 \\
 A4 &  0.90000   &   0.40268   &  0.55524  &  2.38295  &  1.1186   & 1.1179 \\
 A5 &  0.80000   &   0.55626   &  0.76700  &  4.93320  &  0.9557   & 0.9576 \\
 A6 &  0.70000   &   0.65789   &  0.90713  &  7.56798  &  0.7801   & 0.7917 \\
 A7 &  0.60000   &   0.71733   &  0.98909  &  9.86465  &  0.5794   & 0.6176 \\
 A8 &  0.55625   &   0.72524   &  1.00000  &  10.2760  &  0.4749   & 0.5361 \\
  \\
C0 &  1.00000   &   0.00000   &  0.00000  &   0.00000  &   1.0826  & 1.1755    \\
C1 &  0.99792   &   0.55856   &  0.08403  &   0.04561  &   1.0798  & 1.1725 \\
C2 &  0.98333   &   0.15764   &  0.23716  &   0.36682  &   1.0601  & 1.1516 \\
C3 &  0.95000   &   0.27145   &  0.40837  &   1.11303  &   1.0146  & 1.1034 \\
C4 &  0.90000   &   0.38006   &  0.57177  &   2.26285  &   0.9447  & 1.0301 \\
C5 &  0.80000   &   0.52334   &  0.78732  &   4.64841  &   0.7984  & 0.8792 \\
C6 &  0.70000   &   0.61536   &  0.92576  &   7.01965  &   0.6392  & 0.7223  \\
C7 &  0.60000   &   0.66249   &  0.99667  &   8.77258  &   0.4562  & 0.5559 \\
C8 &  0.57656   &   0.66471   &  1.00000  &   8.87526  &   0.4077  & 0.5147 \\
\hline
\end{tabular}
\end{center}
\end{table}
\subsection{Corotating background} \label{sec1}

The equations that describe rapidly and uniformly rotating background
models can be derived by imposing the conditions of stationarity and axi-symmetry
on the Euler-type
equations~(\ref{eq:Euler}) and the Poisson
equation~(\ref{eq:Poisson})~\citep{2002A&A...381..178P,2004MNRAS.347..575Y}. This leads to
\begin{eqnarray}
&&  \tilde \mu _\x + \Phi - \frac{r^2}{2} \sin ^2  \theta  \, \Omega_\x^2  =  C_\x \, , \label{eq:bgmu_x} \\
&& \Phi \left( \mtb{r} \right)  = - G \int_{0}^{\mtb{r}}
\frac{\rho\left(\mtb{r'} \right)}{ | \mtb{r} - \mtb{r'} |} d \mtb{r'} \label{eq:Pois} \, ,
\end{eqnarray}
where $\Omega_\x$ and $C_\x$ are, respectively, the angular velocities
and the integration constants for the neutron and proton fluids.
For corotating background models,
i.e. when $\Omega_{\n}=\Omega_{\p}=\Omega$, in which the two fluids are in
$\beta$-equilibrium and share a common surface, the hydrostatic
equilibrium equation~(\ref{eq:bgmu_x}) becomes
\begin{equation}
\tilde \mu + \Phi - \frac{r^2}{2} \sin ^2 \theta  \, \Omega^2  =  C \, , \label{eq:bg1}
\end{equation}
where $\tilde{\mu} \equiv \tilde{\mu}_{\p} = \tilde{\mu}_{\n}$ is the
background chemical potential and $C\equiv C_\n = C_\p$. When the
system of equations~(\ref{eq:Pois})--(\ref{eq:bg1}) is closed by an
EoS, we can numerically determine a corotating stationary axisymmetric
background via the self-consistent field
method~\citep{1986ApJS...61..479H, 2009MNRAS.396..951P}. The solution is such that the
surface of the star corresponds to the zero chemical potential surface, $\tilde
{\mu} \left( R\left(\theta\right) , \theta \right) =
0$~\citep{2004MNRAS.347..575Y}.

For the EoS~(\ref{eq:EoSA}) and (\ref{eq:EosPR}) we
construct the sequences of rotating models A and C, respectively. The set of models extends from a
non-rotating model up to the mass shedding limit.  In the numerical
code, we re-write the background equation in dimensionless form by
using the gravitational constant $G$, the central mass density
$\rho_0$ and the equatorial radius $R_{eq}$. All  stellar models of
the sequence have the central proton fraction set to $x_\p(0) =
0.1$. By specifying the axis ratio between the polar and equatorial
radius~$R_{p}/R_{eq}$, the iterative numerical routine determines all
 other quantities of an axisymmetric configuration.  The main
properties of the rotating models are given in
Table~\ref{tab:back-models}. From these quantities we can easily
construct stellar models in physical units. For instance for models C,
we can evaluate equation~(\ref{eq:rhomu}) at the centre and obtain:
\begin{equation}
 \gamma_\n k_{\n}^{\ast} = \tilde{\mu}_{0}^{\ast} \left( 1 - x_\p (0) \right) ^{1-\gamma_\n} \, , \qquad \qquad
 \gamma_\p k_{\p}^{\ast} = \tilde{\mu}_{0}^{\ast} x_\p (0) ^{1-\gamma_\p} \, ,  \label{eq:Kph}
\end{equation}
where the asterisk denotes the dimensionless quantities $k^{\ast}_\x =
k_\x / \left(G R_{eq}^2 \rho_0 ^{2-\gamma_n} \right)$ and $
\tilde{\mu}^{\ast} = \tilde{\mu} / \left(G \rho_0 R_{eq}^2
\right)$. Combining equations~(\ref{eq:Kph}) with the dimensionless
mass $M^{\ast} = M / \left(\rho_0 R_{eq}^2 \right)$, we can derive the
equatorial radius of the star,
\begin{equation}
R_{eq} = \left[ \frac{ k_\n \gamma_\n }{G} M^{\gamma_\n-2} \left( 1 - x_\p (0) \right)^{\gamma_\n -1 }
\right]^{1/\left(3 \gamma_\n -4\right)}
\left( M^{\ast \gamma_\n-2}  \mu_0^{\ast} \right)^{-1/\left(3 \gamma_\n -4\right)} \, ,
\end{equation}
where the physical mass $M$ and the EoS parameters can be arbitrarily
chosen. The central mass density, $\rho_0$, and the rotational period, $P$, are determined by the following equations:
\begin{eqnarray}
\rho_0 & = & 2.8 \times 10^{15} \left( M^{\ast} \right)^{ -1} \left( \frac{M}{1.4 M_{\odot}} \right)
                  \left( \frac{ R_{eq} }{10\ \rm{km}}  \right)^{-3} \rm{g \,cm}^{-3}  \, .\\
P & = &0.4596  \, \left( M^{\ast} \right)^{1/2}  \, \left( \Omega^{\ast} \right)  ^{-1}  \left( \frac{M}{1.4 M_{\odot}} \right)^{-1/2}
              \left( \frac{ R_{eq} }{10\ \rm{km}}  \right)^{3/2} \textrm{ms}  \, ,  \label{eq:Tform}
\end{eqnarray}
where $\Omega^{\ast} = \Omega / \sqrt{G \rho_0 }$.

\begin{figure}
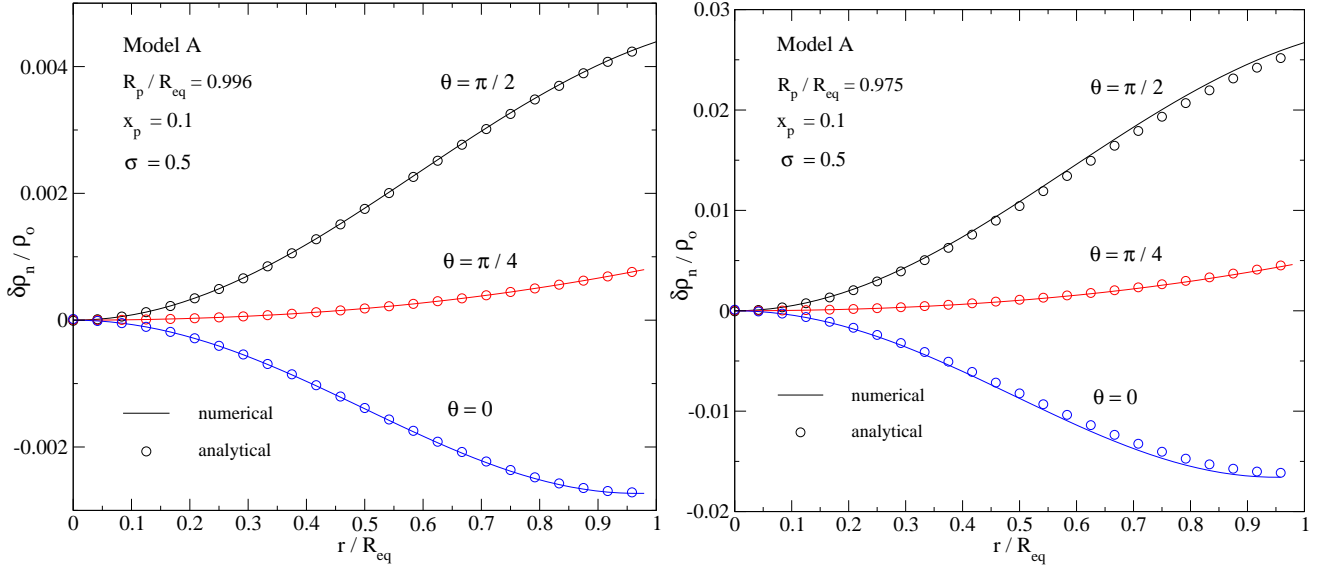

\begin{center}
\includegraphics[height=75mm]{fig1a.eps}
\includegraphics[height=75mm]{fig1b.eps}
\caption{In this figure, we compare our numerical results to the
analytical solution of~\citet{2002A&A...381..178P} for two slowly
rotating stellar models with axis ratio $0.996$ (left panel) and
$0.975$ (right panel). These two background stars are described by the
EoS~(\ref{eq:EoSA}) and have the same proton fraction $x_\p = 0.1$ and symmetry
energy term $\sigma = 0.5$. The non-corotating corrections are
determined by choosing the relative angular velocity~$\left(
\delta \Omega_\n , \delta \Omega_\p \right) = \left( 1 , 0 \right)$ and
imposing the constant central chemical potential
condition~(\ref{eq:Cx}). In the two panels, we show the radial profile
of the perturbed neutron mass density~$\delta \rho _{\n} / \rho_0 $
for three different angular directions, i.e. $\theta = 0, \pi / 4$ and
$\pi / 2$. Our numerical results (solid line) agree very well with the
analytical solution (empty circle) for the slowest rotating model
(left panel). For faster rotating models the slow-rotation
solution is expected to be less accurate. This  is already
evident for the case in the right panel, where the numerical and analytical
solutions start to disagree.
\label{fig:backmodel}}
\end{center}
\end{figure}

\subsection{Non-corotating solutions}
\label{sec:Noncorback}

In a multi-fluid system, like an astrophysical neutron star, the various fluid components
can have different velocities. This is, in fact, an essential element in the
favoured model for pulsar glitches where  the  sudden observed
spin-up is explained as a transfer of angular momentum
between an interior superfluid neutrons and the charged component. In
this model, the momentum transfer is due to the interaction
between the crust and an array of quantised neutron vortices that are
generated by the stellar rotation.  During the magnetically driven spin-down
of a neutron star, these vortices are pinned to the crust
and corotate with the charged components. Therefore, a velocity lag
develops between superfluid neutrons and the crust and an increasing
Magnus force acts on the vortices. When this force becomes
stronger than the pinning force, the vortices should unpin. At this point they
are free to move and can
accelerate the crust, generating a glitch.

Typically, the spin variation observed in a glitch is very small,~$
10^{-9} < \Delta \Omega / \Omega < 10^{-5}$. This means that the effects of a glitch
on the stellar structure is expected to be tiny and can be studied
 perturbatively. The approach developed
by~\cite{2004MNRAS.347..575Y} is particularly appropriate for this
kind of problem, as the non-corotating quantities are considered as
small deviations from a stationary, rapidly corotating configuration.
We have already used these non-corotating corrections as  initial
data for studying the post-glitch dynamics and the associated stellar
oscillations~\citep{2010MNRAS.tmp..554S}. We will now provide further
details about the method.

Adopting the~\cite{2004MNRAS.347..575Y} approach, we expand
equations~(\ref{eq:bgmu_x})-(\ref{eq:Pois})  up to the first order in
\begin{displaymath}
\left( \Omega_\n -
\Omega_\p \right) / \left( |\Omega_\n| + |\Omega_\p| \right)\ .
\end{displaymath}
Thus, we have
\begin{eqnarray}
\Omega_\x & = & \Omega_\crt \left( 1 + \delta \Omega_\x \right) \, , \\
\rho _ \x & = & \rho_\crt + \delta \rho _\x \, , \\
\mu _ \x  & = &  \mu_\crt + \delta  \mu _\x \, , \\
\Phi      & = & \Phi_\crt + \delta \Phi \, ,
\end{eqnarray}
where the subscript ``c'' denotes the corotating values. Note that by
definition $\delta \Omega_\x$ represents the relative deviation of the
x fluid angular velocity with respect to the corotating background,
i.e. $\delta \Omega_\x = \left(  \Omega_\x - \Omega_{\crt} \right) / \Omega_{\crt}$.
For the non-corotating
corrections, Equations~(\ref{eq:bgmu_x})--(\ref{eq:Pois}) become
\begin{eqnarray}
&& \delta \tilde \mu _\x + \delta \Phi - r^2 \sin ^2 \theta \,
\Omega_\crt ^2  \, \delta \Omega_\x = \delta C_\x \, ,
\label{eq:muxNC} \\
&& \delta \Phi \left( \mtb{r} \right)  = - G \int_{0}^{\mtb{r}}
\frac{ \delta \rho\left(\mtb{r'} \right)}{ | \mtb{r} - \mtb{r'} |} d \mtb{r'} \, . \label{eq:PoisNC}
\end{eqnarray}
The system of equations is closed by
\begin{equation}
\delta \tilde \mu _{\x} = \left. \frac{ \partial \tilde \mu_{\x} }{ \partial \rho_{\p} } \right|_{\rho_\n} \delta \rho_{\p}
                        + \left. \frac{ \partial \tilde \mu_{\x} }{ \partial \rho_{\n} } \right|_{\rho_\p} \delta \rho_{\n} \, ,
\label{eq:tildemu}
\end{equation}
that relates the mass density and the
chemical potential perturbations for co-rotating backgrounds.

Non-corotating solutions can be constructed with either fixed
central chemical potential or total mass.  For the
first class of models, we can impose the
condition~$\left. \delta \tilde \mu_\x \right|_{r=0} = 0$ at the star's centre, and
determine the integration constant $\delta C_{\x}$ from
equation~(\ref{eq:muxNC})~\citep{2004MNRAS.347..575Y}:
\begin{equation}
\delta C_\p = \delta C_\n = \left. \delta \Phi \right|_{r=0} \,
. \label{eq:Cx}
\end{equation}
For solutions with constant mass, we impose a constraint on the mass of each
fluid component, i.e.
\begin{equation}
\delta M_\x \equiv \int d \mtb{r} \, \delta \rho_\x = 0 \, .
\end{equation}
In equation~(\ref{eq:muxNC}), we can replace the chemical potential by
the mass density perturbation using equation~(\ref{eq:tildemu}), and
integrate over the star's volume $V$. The integration constant $\delta
C_{\x}$ is then given by the following expression:
\begin{equation}
\delta C_\x \int d\mtb{r} \rho_\x^{2-\gamma_\x} = \int d\mtb{r} \rho_\x^{2-\gamma_\x}
\left(  \delta \Phi -r^2 \sin ^2 \theta \Omega_c \delta \Omega_x \right) \,
. \label{eq:CxFM}
\end{equation}
For the EoS~(\ref{eq:EoSA}), the adiabatic index is $\gamma_\x = 2$
and the boundary condition~(\ref{eq:CxFM}) therefore reduces to:
\begin{equation}
\delta C_\x  = \frac{1}{V} \int d\mtb{r} \left(  \delta \Phi -r^2 \sin ^2\theta \Omega_c \delta \Omega_x \right) \,
. \label{eq:CxFMA}
\end{equation}

\begin{figure}
\begin{center}
\includegraphics[width=85mm]{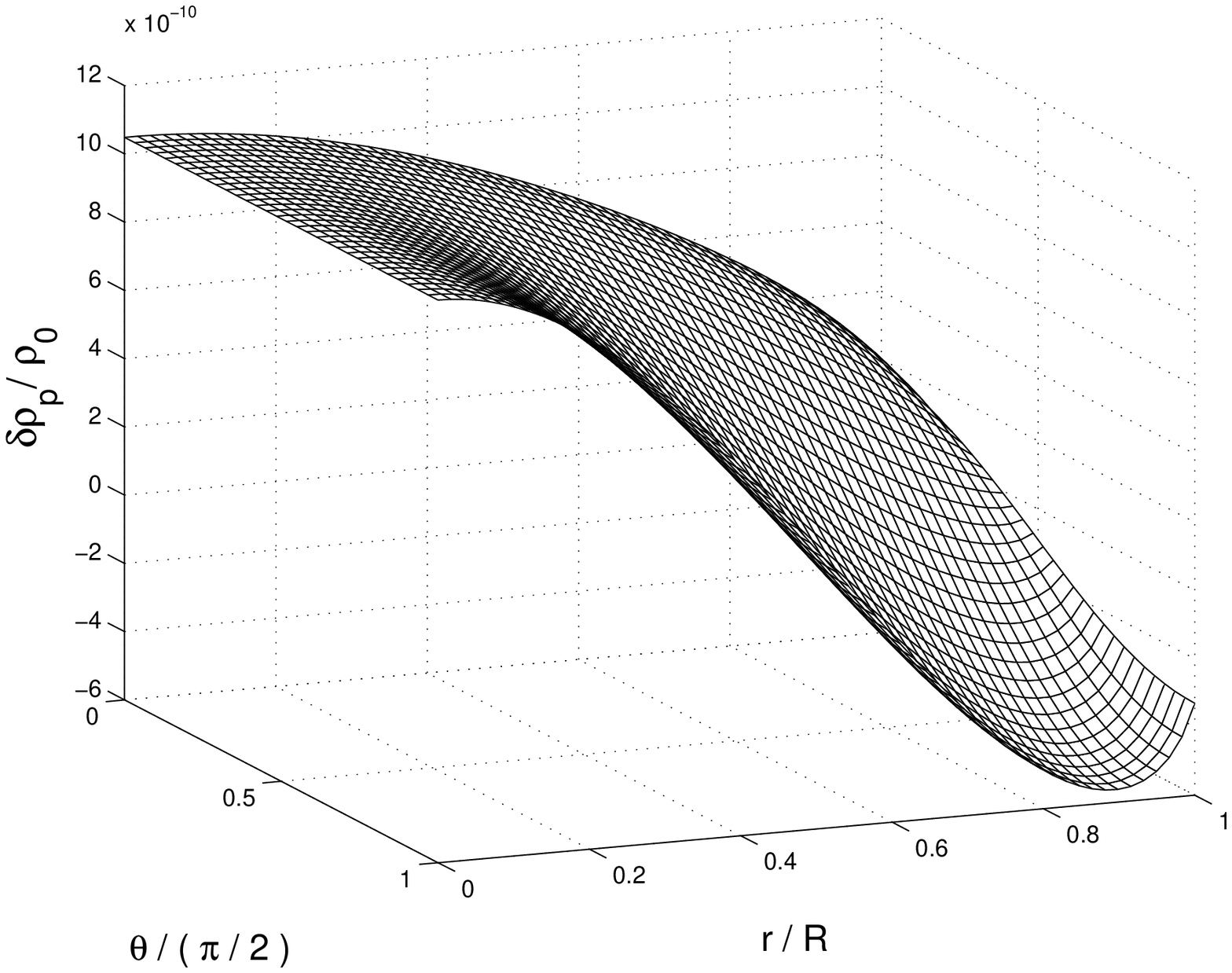}
\includegraphics[width=85mm]{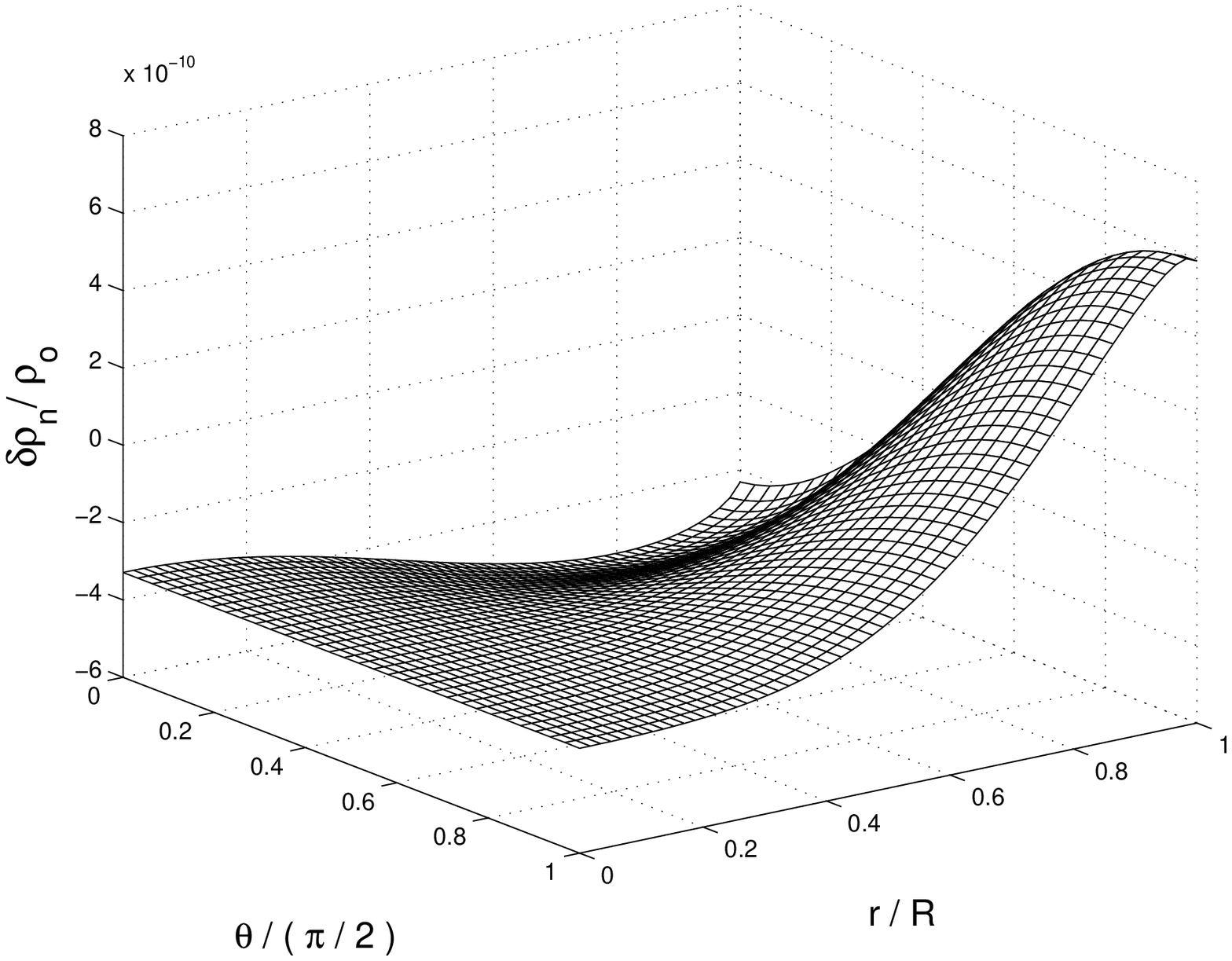}
\caption{\label{eq:gl-IDrho} This figure displays, for model C2, the
  non-corotating solutions of the proton and neutron mass density
  $\delta \rho_\x / \rho_{0}$, in the left and right
  panel, respectively. The results correspond to constant mass solutions with
  parameters~$\delta \Omega _\p = - 10^{-6}$ and $\delta \Omega_\n =
  7.74\times 10^{-8}$. These solutions were used as initial
  conditions by~\citet{2010MNRAS.tmp..554S} for studying glitch
  hydrodynamics.}
\end{center}
\end{figure}

The system of equations~(\ref{eq:muxNC})--(\ref{eq:PoisNC}) can be
solved iteratively. First of all, for a given EoS we determine the
co-rotating background with the self-consistent field method
of~\citet{1986ApJS...61..479H}, where we specify the axis ratio of the
star. Secondly, we choose the relative angular velocity $\delta \Omega_\x$ of each fluid
component. The iteration algorithm then proceeds as
follows: \emph{i)} we solve the perturbed Poisson
equation~(\ref{eq:PoisNC}) for an initial guess of the perturbed mass
density $\delta \rho$, \emph{ii)} we get the integration constant
$\delta C_x$ imposing either the condition~(\ref{eq:Cx})
or~(\ref{eq:CxFM}), \emph{iii)} we determine the chemical potential
$\delta \tilde \mu _\x$ from equation~(\ref{eq:muxNC}) and then the
new mass density $\delta \rho$ from the EoS. This procedure is
iterated until the difference between the quantities is smaller than a
prescribed error.

An important property of this linear perturbation approach is that we
can construct two independent solutions to
equations~(\ref{eq:muxNC})--(\ref{eq:PoisNC}),  respectively
corresponding to $\left( \delta \Omega_\n , \delta \Omega_\p \right) =
\left( 1 , 0 \right)$ and $\left( \delta \Omega_\n , \delta \Omega_\p
\right) = \left( 0 , 1 \right)$. Since the problem is linear, any non-corotating configuration can
be obtained as a linear combination of these two solutions.

We have tested our code against the analytical solution for the  EoS~(\ref{eq:EoSA}) determined
by~\cite{2002A&A...381..178P} in the slow-rotation approximation. We select two slowly
rotating models with axis ratio $0.996$ and $0.975$,
respectively. The models have the same proton fraction and symmetry
energy term, i.e. $x_\p = 0.1$ and $\sigma = 0.5$. The non-corotating
corrections correspond to a relative angular velocity $\left( \delta
\Omega_\n , \delta \Omega_\p \right) = \left( 1 , 0 \right)$ with
constant central chemical potential, c.f.~(\ref{eq:Cx}). In
Fig.~\ref{fig:backmodel}, we show the radial profile of the perturbed
neutron mass density~$\delta \rho _{\n} / \rho_{0}$ for the three
angles $\theta = 0, \pi / 4$ and $\pi / 2$, respectively. In the
slowest rotating model, the agreement between the numerical and the analytical solutions is evident. In the
second model, with axis ratio~$0.975$ the two solutions begin to
differ, as expected. The slow-rotation solution becomes less accurate
as the star's rotation increases. The same behaviour is found for
non-corotating solutions with constant mass, i.e. when~$\delta M_\x = 0$.
This comparison gives us confidence in our numerically generated background models.

\begin{figure}
\begin{center}
\includegraphics[width=85mm]{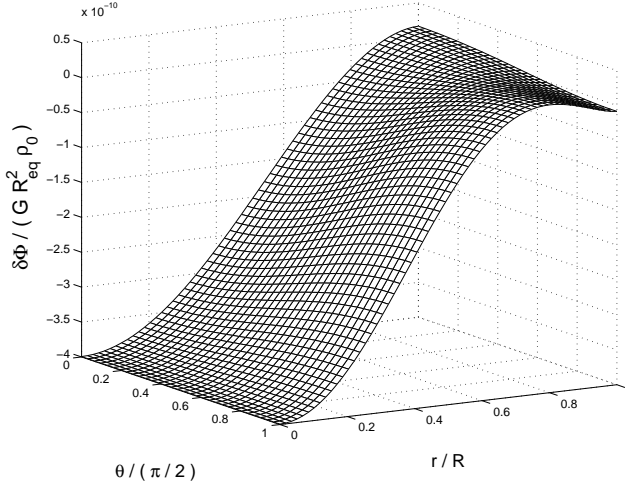}
\caption{\label{eq:gl-IDPhi} We show, for the C2 model and the
non-corotating configuration from Fig.~\ref{eq:gl-IDrho}, the result
for the dimensionless gravitational potential~$\delta \Phi / \left( G
R_{eq}^{2} \rho_{0}\right)$.}
\end{center}
\end{figure}

For the sequence of constant mass models, we show in
Figs.~\ref{eq:gl-IDrho} and~\ref{eq:gl-IDPhi} the non-corotating mass
density pertubations~$\delta \rho_\x / \rho_{0}$ and the gravitational
potential perturbation~$\delta \Phi$ for the C2 model.  These are
solutions to equations~(\ref{eq:muxNC})--(\ref{eq:PoisNC}) with~$\delta
\Omega _\p = - 10^{-6}$ and $\delta \Omega_\n = 7.74\times 10^{-8}$,
which were used as initial conditions for the glitch
simulations discussed by~\cite{2010MNRAS.tmp..554S}.

\section{Perturbation Equations} \label{sec:PertEqs}

The dynamics of a superfluid neutron star can be studied by
linearizing the system of differential
equations~(\ref{eq:Mcon})--(\ref{eq:Poisson}). In the inertial frame,
the Eulerian perturbation equations are given by
\begin{eqnarray}
\partial_t \left( \delta \mtb{v}_{\x} + \veps_\x \delta
  \mtb{w}_{\y\x} \right) + \Omega \, \partial_{\phi} \delta \mtb{v}_{\x} & = & - \nabla \delta \tilde{\mu}_\x - \nabla
\delta \Phi - 2 \mtb{\Omega} \times \delta \mtb{v}_{\x} + \frac{\delta
  \mtb{f^{\x}}}{\rho_{\x}} \, ,
             \label{eq:dVxdt} \\
\left( \partial_t  + \Omega \, \partial_{\phi} \right) \, \delta \rho_\x & = & -  \nabla \cdot  \left( \rho_\x \delta \mtb{v}_\x \right)   \, ,        \label{eq:drhoxdt} \\
\nabla ^2  \delta \Phi   & = &  4 \pi G \, \delta \rho     \, ,        \label{eq:dPhi-poiss}
\end{eqnarray}
where $\phi$ is the azimuthal angle associated with the rotational motion, and
the perturbed mutual friction force is (in the case of a co-rotating background)
\begin{equation}
\delta \mtb{f}^{\x} = 2 \rho_\n \mathcal{B}' \mtb{\Omega} \times
\delta \mtb{w}_{\x\y} + 2 \rho_\n \mathcal{B} \, \hat{\mtb{\Omega}}
\times \mtb{\Omega} \times \delta \mtb{w}_{\x\y} \, .
\end{equation}
The chemical potential perturbations can be expressed  in
terms of the mass density perturbations using
equation~(\ref{eq:tildemu}).

In order to solve numerically
equations~(\ref{eq:dVxdt})--(\ref{eq:drhoxdt}) we use the conjugate
momentum perturbations $\delta \mtb{p}_\x$ as dynamical variables. These are given by
\begin{eqnarray}
\delta \mtb{p}_\n  & = & \left( 1 - \veps_\n \right) \delta \mtb{v}_\n + \veps_\n \delta \mtb{v}_\p  \, , \label{eq:def-Dn} \\
\delta \mtb{p}_\p  & = & \veps_\p \delta \mtb{v}_\n +  \left( 1 - \veps_\p \right) \delta \mtb{v}_\p \, , \label{eq:def-Dp}
\end{eqnarray}
where we recall that $\rho_\p \veps_\p =\rho_\n\veps_\p$.
By inverting these relations we can determine the velocity fields at any time step,
\begin{eqnarray}
\delta \mtb{v}_\n  & = & \frac{ \left(1 - \veps_\p \right) \delta \mtb{p}_\n - \veps_\n \delta \mtb{p}_\p }{1-\bar \veps } \, ,\label{eq:def-Vn} \\
\delta \mtb{v}_\p  & = & \frac{ - \veps_\p \delta \mtb{p}_\n +  \left( 1 - \veps_\n \right) \delta \mtb{p}_\p }{1-\bar \veps } \, , \label{eq:def-Vp}
\end{eqnarray}
where $\bar \veps \equiv \veps_\n + \veps_\p = \veps_\n  / x_\p $.

The time evolution of the non-axisymmetric perturbation equations is a
three-dimensional problem in space. However, linear perturbations on
an axisymmetric background can be expanded in terms of a set of basis
functions $\left( \cos m \phi \, ,\sin m \phi \right)$, where $m$ is
the azimuthal harmonic index~\citep{1980MNRAS.190...43P}. The mass
density perturbations as well as the other perturbation quantities
then take the following
form~\citep{2002MNRAS.334..933J,2009MNRAS.396..951P}
\begin{equation}
\delta \rho \left( t,r,\theta,\phi \right) = \sum_{m=0}^{m=\infty}
               \left[ \delta \rho_{ m}^{+} \left( t,r,\theta\right)
               \cos m \phi + \delta \rho_{m}^{-} \left(
               t,r,\theta\right) \sin m \phi \right] \, .
               \label{eq:drhoexp}
\end{equation}
With this Fourier expansion the perturbation equations decouple with
respect to $m$ and the problem becomes two-dimensional. In particular,
for the axisymmetric case ($m=0$) only the $\delta \rho_{0}^{+}$
component survives.

\subsection{Boundary Conditions} \label{sec:BC}

In this work, we study axisymmetric ($m=0$) and non-axisymmetric
oscillations ($m=2$) of a superfluid neutron star with equatorial and
rotational axis symmetry. The numerical domain extends over the region
$0 \leq r/R(\theta) \leq 1 $ and $ 0\leq \theta \leq \pi / 2$, and we
need to impose boundary conditions at the surface, origin, rotational
axis and equator.

We first discuss the boundary conditions at the origin~($r=0$) and the
rotational axis~($\theta=0$), where the perturbation equations must be
regular.  Let us denote by $\delta \psi$ a general scalar
perturbation, such as the mass density~$\delta \rho_\x$, the chemical
potential~$ \delta \tilde{\mu}_\x$ and the gravitational potential
$\delta \Phi$.  For axi-symmetric and non-axisymmetric oscillations,
we have to impose the following conditions,  respectively :
\begin{eqnarray}
\left. \frac{ \partial \delta \psi}{\partial r} \right|_{r=0} & = &
\left. \frac{  \partial \delta \psi}{\partial \theta} \right|_{\theta=0} = 0 \quad \mbox{for} \quad m=0 \, ,\\
\left. \delta \psi \right|_{r=0} & = & \hspace{2.8mm} \left. \delta \psi \right|_{\theta=0}  = 0 \quad \mbox{for} \quad m=2 \,  .
\end{eqnarray}
For the velocity fields~$\delta \mtb{v}_{\x}$, we impose that there
must be no mass flux across the origin~($r=0$) for both axisymmetric
and non-axisymmetric perturbations:
\begin{equation}
\left. \delta v^{r}_{\x} \right|_{r=0} = 0 \, .
\end{equation}
At the rotational axis~($\theta = 0$), we impose the following
conditions:
\begin{eqnarray}
\frac{ \partial \delta v^{r}_{\x}}{\partial \theta} & = & \delta v^{\theta}_{\x} =
 \delta v^{\phi}_{\x} = 0  \quad \mbox{for} \quad m=0  \, , \\
\delta v^{r}_{\x}  & = & \delta v^{\theta}_{\x} =
 \delta v^{\phi}_{\x} = 0 \quad \mbox{for} \quad m=2  \, .
\end{eqnarray}

At the equator~($\theta = \pi/2$), the reflection symmetry divides the
perturbations  into two sets with opposite
parity~\citep{2009MNRAS.394..730P}. In the Type~I parity class, the
scalar perturbations $\delta \psi$ and the velocity satisfy the
following conditions:
\begin{equation}
\frac{ \partial \delta \psi }{\partial \theta} = \frac{ \partial \delta v^{r}_{\x}}{\partial \theta} = \delta v^{\theta}_{\x} =
 \frac{ \partial \delta v^{\phi}_{\x}}{\partial \theta} = 0 \, .
\end{equation}
Meanwhile, the Type II class is such that:
\begin{equation}
 \delta \psi  =  \delta v^{r}_{\x} = \frac{ \partial \delta v^{\theta}_{\x}}{\partial \theta}
= \delta v^{\phi}_{\x} = 0 \, .
\end{equation}

The outer layers of a mature neutron star form an elastic crust made up of nuclei.
The crust is an important aspect that is yet to be implemented in our numerical
model (although we are making progress on it). Our current model is
simplified, in the sense that  we assume that superfluid neutrons and
protons are present throughout the stellar volume.  We then impose the standard boundary
condition of a free surface, i.e.  require that  the Lagrangian perturbation of the
individual chemical potentials vanish
at the surface, i.e.
\begin{equation}
\Delta \tilde {\mu} _\x = \delta \tilde {\mu} _\x + \mtb{\xi} _\x
\cdot \nabla \tilde{\mu}_{\crt} = 0 \, . \label{eq:bcsurf}
\end{equation}
The vector field $\mtb{\xi} _\x$ is the Lagrangian displacement of the
x-fluid component~\citep*{2004MNRAS.355..918A}. The value of the
perturbed chemical potential $\delta \tilde {\mu} _\x$ at the surface
is  determined from equation~(\ref{eq:bcsurf}) at each time step.

\section{Gravitational-wave Extraction} \label{sec5}

In order to study the gravitational-wave signal emitted by pulsating
superfluid neutron stars, we have implemented the quadrupole
formula for both axisymmetric and non-axisymmetric oscillations.  We will now discuss this
implementation, in particular, the momentum and stress formula that we use to
improve the
numerical gravitational-wave extraction.

The gravitational-wave strain  can be determined using
 the quadrupole formula~\citep{1980RvMP...52..299T}:
\begin{equation}
h_{ij}^{2m} = \frac{G}{c^4} \frac{1} {r} \sum_{m=-l}^{l} \frac{d^2
\mathcal{I}}{dt^2}^{2m} \, T_{ij} ^{E2,2m} \, ,  \label{eq:hp22}
\end{equation}
where $T_{ij} ^{E2,2m}$ is the pure spin tensor harmonic which has
``electric-type'' parity, i.e. $\left( -1
\right)^{l}$~\citep{1980RvMP...52..299T}. In this work, we focus only
on the $m=0$ and $m=2$ pulsations. In the orthonormal basis of
spherical coordinates, the components of the $(l,m)=(2,0)$ and
$(l,m)=(2,2)$  spin tensor harmonics are, respectively, given by
\begin{eqnarray}
 T_{\theta \theta} ^{E2,20} & = & \frac{1}{8} \sqrt{ \frac{ 15}{\pi }} \sin^2 \theta  \, ,
\end{eqnarray}
and
\begin{eqnarray}
 T_{\theta \theta} ^{E2,22} & = & T_{\phi \phi} ^{E2,22} = \frac{1}{16} \sqrt{ \frac{ 10}{\pi }} \left( 1 + \cos ^2 \theta  \right) e^{2i\phi}  \, , \\
 T_{\theta \phi} ^{E2,22} & = & \frac{i}{16} \sqrt{ \frac{ 10}{\pi }} \cos \theta  \, e^{2i\phi}  \, .
\end{eqnarray}
The quantity $\mathcal{I}^{2m}$ is the quadrupole moment, in the case of a two-fluid star defined by;
\begin{equation}
\mathcal{I}^{2m} = \frac{16 \pi }{15} \sqrt{3}   \int d\mtb{r} \, \delta \rho \, r^2 \, Y_{2m}^{\ast}
=   \frac{16 \pi }{15} \sqrt{3}  \int d\mtb{r} \, \left(  \delta \rho_\n +  \delta \rho_\p \right)  r^2 \, Y_{2m}^{\ast}  \, , \\
\end{equation}
where the spherical harmonics $Y_{2m}$ for the $m=0$ and $m=2$ cases
are given by
\begin{eqnarray}
Y_{20} & = & \frac{1}{4} \sqrt{ \frac{5}{\pi} } \left( 3 \cos ^2
\theta -1 \right) = \frac{1}{2} \sqrt{ \frac{5}{\pi} }
P_{20}\left(\cos \theta \right) \, , \\ Y_{22} & = & \frac{1}{4}
\sqrt{ \frac{15}{2\pi} } \sin^2 \theta \, e^{2i\phi}\, .
\end{eqnarray}
where $P_{20}\left(\cos \theta \right)$ is the Legendre polynomial.

It is well-known that, the numerical calculation of the second
order time derivative of the quadrupole moment in equation~(\ref{eq:hp22})
could lead to
inaccurate results~\citep{1990ApJ...351..588F}.  However, the accuracy
of the gravitational-wave extraction can be improved by transforming
equation~(\ref{eq:hp22}) into either the perturbed momentum formula,
with a first order time derivative, or the perturbed stress
formula, where the time derivatives are
absent~\citep{1990ApJ...351..588F}.  In this work, we use both these prescriptions
 in order to check the wave extraction accuracy.

For axisymmetric oscillations, $m=0$, the gravitational strain can be
written as follows:
\begin{equation}
h^{20} = \frac{G}{c^4} \frac{\sin^2 \theta} {r} \sum_{\x}
A_\x^{20} \, ,  \label{eq:hp3}
\end{equation}
where the quantity $A_\x^{20}$ is defined by
\begin{equation}
A^{20}_\x \equiv  \frac{d^2 }{dt^2}  \int d\mtb{r}  \,  \delta \rho_\x \, r^2 \, P^{20}
 \, .
\label{eq:A20-dens}
\end{equation}
We can reduce the order of the time derivative by using the method
developed by~\citet{1990ApJ...351..588F}, and obtain the perturbed
momentum formula:
\begin{equation}
A^{20}_\x \equiv 2 \frac{d }{dt} \int d\mtb{r} \, r \rho_\x \left(
\delta v_\x^{r} \, P^{20} + \frac{ \delta v_\x^{\theta} }{2} \frac{
\partial P}{\partial \theta}^{\hspace{-0.05cm}20} \right) \, ,
\label{eq:A20-mom}
\end{equation}
and the perturbed stress formula:
\begin{eqnarray}
A^{20}_\x & \equiv & 2  \int   d\mtb{r} \left\{ - \Omega r \sin \theta \, \rho_\x \, \delta
v_\x^{\phi} - \frac{1}{2} \left( \Omega r \sin \theta \right)^2 \delta
\rho_\x \nn \right. \\ & + & \left. \frac{1}{4\pi} \left[ \nabla_{r}
\Phi \nabla_{r} \delta \Phi P^{20} \hspace{-0.1cm} \left(\cos \theta
\right) + \nabla_{\theta} \Phi \nabla_{\theta} \delta \Phi \, P^{20}
\hspace{-0.1cm} \left( \sin \theta \right) + \frac{1}{2} \left(
\nabla_{r} \Phi \nabla_{\theta} \delta \Phi + \nabla_{\theta} \Phi
\nabla_{r} \delta \Phi \right) \partial_{\theta} P
^{\hspace{-0.01cm}20} \hspace{-0.1cm} \left(\cos \theta \right)
\right] \right\} \, ,
\label{eq:A20-stres}
\end{eqnarray}
where the gradient components $\nabla_{i}$ in eqaution~(\ref{eq:A20-stres}) are determined in the
orthonormal spherical basis, i.e. $\nabla = \left( \partial_{r},
\frac{1}{r} \partial_{\theta}, \frac{1}{r \sin \theta} \partial_{\phi}
\right)$. At the end of the day, the quantity $A^{20}_\x$ in the strain
equation~(\ref{eq:hp3}) can be determined from either of the three
equations~(\ref{eq:A20-dens})--(\ref{eq:A20-stres}).

\begin{figure}
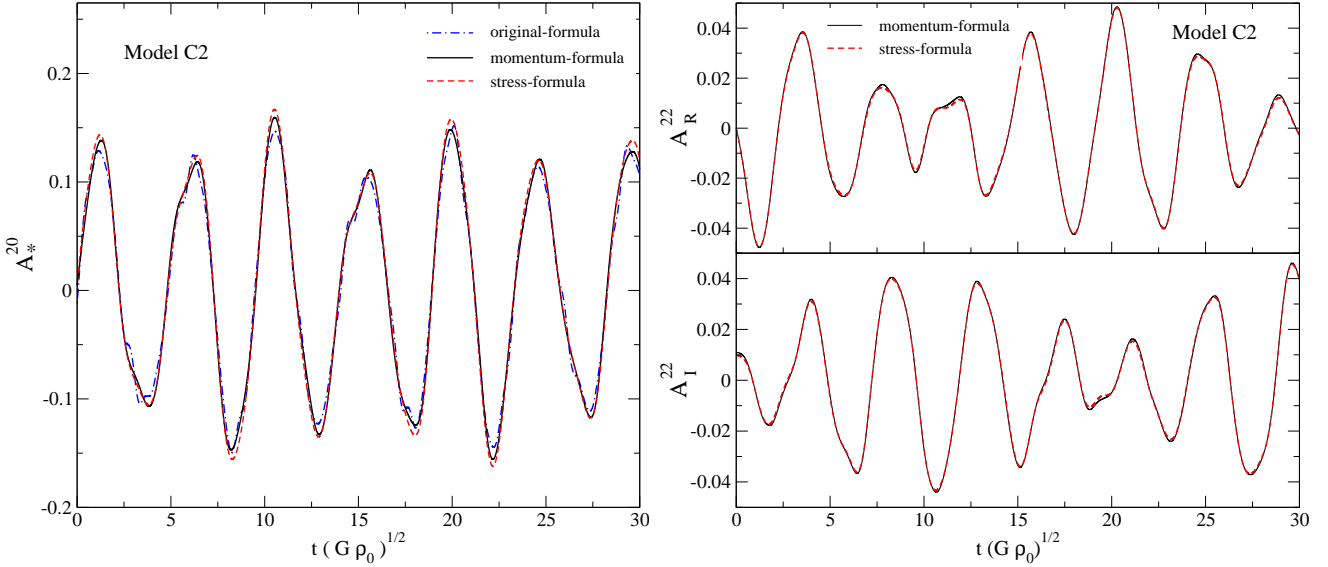

\begin{center}
\includegraphics[height=75mm]{fig4a.eps}
\hspace{0.2mm}\includegraphics[height=75mm]{fig4b.eps}
\caption{ We compare the gravitational-wave extraction results for
  axisymmetric and non-axisymmetric oscillations. The signal is
  generated by perturbing the stellar model C2, and the illustrated
  quantities are dimensionless.  The left panel shows the
  dimensionless code quantity $A^{20}_{\ast}$ determined from three
  equivalent equations, respectively, the second time derivative of the
  quadrupole moment (dot-dashed line), the momentum-formula
  (solid-line) and the stress formula (dashed-line). In the right
  panel, we show the waveform of the $m=2$ non-axisymmetric
  oscillations for the perturbed C2 model. The upper and lower right
  panels displays respectively the real part $A^{22}_{R}$ and the imaginary
  part $A^{22}_{I}$  of the dimensionless quantity
  $A^{22}_{\ast}$ determined by the code. We compare the signal extraction to the
  momentum-formula (solid-line) and the stress formula
  (dashed-line). \label{fig-gwcomp}}
\end{center}
\end{figure}

For non-axisymmetric oscillations with $l=m=2$, the two
independent polarizations of the strain can be written as follows:
\begin{equation}
h_{\theta \theta}^{22} - i h_{\theta \phi}^{22} = h^{22} {}_{-2}Y^{22} \, ,
\end{equation}
where ${}_{-2}Y^{22}$ is the $s=-2$ spin-weighted spherical harmonics,
\begin{equation}
{}_{-2}Y^{22} = \frac{1}{8} \sqrt{\frac{5}{\pi}} \left( 1 + \cos \theta \right) ^2 e^{2i\phi} \, ,
\end{equation}
and we have defined the  quantity
\begin{equation}
h^{22} \equiv \frac{G}{c^4} \frac{8 \pi}{15} \frac{\sqrt{3}}{r} \frac{d^2 }{dt^2}
\int d\mtb{r} \, \delta \rho \, r^2 Y_{22}^{\ast}  \, . \label{eq:h22b}
\end{equation}
We can then re-write equation~(\ref{eq:h22b}) as
follows:
\begin{equation}
h^{22} = \frac{G}{c^4} \frac{8 \pi}{15} \frac{\sqrt{3}}{r} \sum_{\x}
A_\x^{22} \, , \label{eq:hp3a}
\end{equation}
where
\begin{equation}
A^{22}_\x \equiv \frac{d^2 }{dt^2}
\int d\mtb{r} \, \delta \rho_{\x} \, r^2 Y_{22}^{\ast} \, . \label{eq:A22quad}
\end{equation}
In equation~(\ref{eq:A22quad}), the order of the time derivatives can
be reduced by using the equations of motion~(see Appendix~\ref{sec:gwextr}
for more details). This leads to the following expression:
\begin{equation}
A^{22}_\x \equiv  2  \frac{d}{dt} \int d\mtb{r} \left\{ \rho_{\x} r \left[ \left( \delta v^{r}_{\x} - i \frac{\delta v^{\phi}_{\x} }{\sin \theta} \right) Y_{22}^{\ast}
+ \frac{\delta v_{\x} ^{\theta} }{2} Y_{22,\theta}^{\ast} \right] - i \Omega r^2 \delta \rho_{\x} \,  Y_{22}^{\ast} \right\} \, .
\label{eq:A22-mom}
\end{equation}
For linear perturbations on a corotating background, we can further
 transform equation~(\ref{eq:A22-mom}) into the following expression:
\begin{eqnarray}
A^{22}_\x & \equiv &  \frac{1}{2} \sqrt{\frac{15}{2\pi}}  \int d\mtb{r} \left\{
  - 2 \rho_{\x} \Omega r \sin \theta \left[ \delta v^{\phi}_{\x}
+ i \left( \sin \theta  \delta v^{r}_{\x} + \cos \theta \delta v^{\theta}_{x} \right) \right]
- \left( \Omega r \sin \theta \right) ^2 \delta \rho_{\x}
\nn \right. \\
& + & \left.
 \frac{1}{4\pi} \left[  \sin ^2\theta   \, \nabla_{r} \Phi \nabla_{r} \delta \Phi
+ \cos ^2 \theta  \, \nabla_{\theta} \Phi \nabla_{\theta} \delta \Phi
+ \sin\theta \cos \theta \left( \nabla_{r} \Phi \nabla_{\theta} \delta \Phi
+ \nabla_{\theta} \Phi \nabla_{r} \delta \Phi \right)
\nn \right. \right. \\
& - & \left. \left.
i \left( \sin \theta \nabla_{r} \Phi + \cos \theta \nabla_{\theta} \Phi \right) \nabla_{\phi} \delta \Phi
\right]   \right\}  \, ,
\label{eq:A22-stress}
\end{eqnarray}
where the time derivatives are absent.  In
equations~(\ref{eq:A22-mom}) and~(\ref{eq:A22-stress}), the
perturbations are determined in the inertial frame.

The energy radiated as gravitational waves is determined by the
following equation~\citep{1980RvMP...52..299T}:
\begin{equation}
 E_{rad}^{2m} = \frac{1}{32\pi} \frac{G}{c^5} \int_{-\infty}^{\infty}
 \left| \frac{d^3 \mathcal{I}}{dt^3}^{\hspace{-0.04cm}2m} \right| ^{2}
 dt \, .
\label{eq:GWEn}
\end{equation}
By using  Parseval's Theorem we can write equation~(\ref{eq:GWEn})
for the $(2,0)$ and $(2,2)$ components as follows:
\begin{eqnarray}
 E_{rad}^{20} & = & \int_{0}^{\infty} \frac{ d E}{d \nu}^{\hspace{-0.2mm} 20}  d \nu
= \frac{16}{15}\pi^2 \frac{G}{c^5}  \int_{0}^{\infty} \nu^2 \left| \hat{A}^{20} \right| ^2 d \nu \, , \\
 E_{rad}^{22} & = & \int_{0}^{\infty} \frac{ d E}{d \nu}^{\hspace{-0.2mm} 22}  d \nu
= \frac{64}{75}\pi^3 \frac{G}{c^5}  \int_{0}^{\infty} \nu^2 \left| \hat{A}^{22} \right| ^2 d \nu \, ,
\end{eqnarray}
where $A^{2m} = A^{2m}_\n + A^{2m}_\p $, and $\hat{A}^{2m}$ is its
Fourier transformation.

The characteristic strain of the gravitational-wave signal is then given
by~\citep{1998PhRvD..57.4535F}:
\begin{equation}
 h_c \left(\nu \right) \equiv  \sqrt{\frac{{2G}}{\pi^2 c^3}} \frac{1}{d} \sqrt{ \frac{  d E}{d \nu} }
\label{eq:hcdef} \, ,
\end{equation}
where $d$ is the source distance.
The strains $h^{20}$ and $h^{22}$ are related to the
dimensionless quantities $A _{\ast} ^{20}$ and $A _{\ast} ^{22}$ used
in the numerical code by the following expressions:
\begin{eqnarray}
h^{20} & = & 1.414 \times 10^{-17} A _{\ast} ^{20} \left( M^{\ast}
 \right)^{-2} \left( \frac{M}{1.4 M_{\odot}} \right) ^2 \left( \frac{
 R_{eq} }{ 10 \, \rm{km} } \right) ^{-1} \left( \frac{1\ \rm{kpc}}{d}
 \right) \sin ^2 \hspace{-0.5mm} \theta  \, ,   \label{eq:st-unit} \\
h^{22} & = & 4.109 \times 10^{-17} A _{\ast} ^{22} \left( M^{\ast}
 \right)^{-2} \left( \frac{M}{1.4 M_{\odot}} \right) ^2 \left( \frac{
 R_{eq} }{ 10 \, \rm{km} } \right) ^{-1} \left( \frac{1\ \rm{kpc}}{d}
 \right) \, .  \label{eq:st2-unit}
\end{eqnarray}
  Similar relations provide the characteristic strain
\begin{eqnarray}
h_{c}^{20} & = & 1.513 \times 10^{-18} |\hat A^{20}_{\ast}| \left( M^{\ast}
 \right)^{-3/2} \left( \frac{M}{1.4 M_{\odot}} \right) ^{3/2} \left( \frac{
 R_{eq} }{ 10 \, \rm{km} } \right) ^{1/2} \left( \frac{1\ \rm{kpc}}{d}
 \right) \left( \frac{ \nu }{ 1 \rm{kHz} }
 \right)  \, ,  \label{eq:stc-unit} \\
h_{c}^{22} & = & 2.398 \times 10^{-18} |\hat A^{22}_{\ast}| \left( M^{\ast}
 \right)^{-3/2} \left( \frac{M}{1.4 M_{\odot}} \right) ^{3/2} \left( \frac{
 R_{eq} }{ 10 \, \rm{km} } \right) ^{1/2} \left( \frac{1\ \rm{kpc}}{d}
 \right) \left( \frac{ \nu }{ 1 \rm{kHz} }  \right) \, .
\end{eqnarray}

As a first test of the numerical implementation,
we compare the gravitational-wave extraction formulae
for the axisymmetric and non-axisymmetric oscillations. We evolve the
C2 model with a density perturbation and extract the signal using
equations~(\ref{eq:A20-dens})--(\ref{eq:A20-stres}) for the $m=0$
pulsations, and (\ref{eq:A22-mom})--(\ref{eq:A22-stress}) for the
$m=2$ oscillations. Typical results are shown in Figure~\ref{fig-gwcomp}. We generally
find good
agreement between the different numerical results, although we note that
(as expected) the momentum and stress formulae produce a
smoother signal than the ``raw'' quadrupole formula~(\ref{eq:A20-dens}).

As an additional test, we have used  the relativistic numerical code developed
by~\cite{Nagar_pr} and \cite{2007PhRvD..75h4038P} to test the results
of the gravitational-wave extraction routine.  In the relativistic case,
the linear perturbations of non-rotating relativistic stars were evolved and
 the signal was extracted using  the Zerilli
function~\citep{Zerilli:1970fj}. From the Newtonian approach used in
the current work, it is evident that we cannot accurately reproduce
the relativistic results. However, we can establish that our
calculations provide a good estimate of the amplitude of the
gravitational-wave strain. To this end, we consider a star with mass $M=1.4M_\odot$
and radius $R=14~\rm{km}$, and evolve the relativistic code with an initial
enthalpy  perturbation, which produces an averaged pulsational
kinetic energy of $\langle E_{k} \rangle \simeq 5.62 \times 10^{-9}
M_{\odot} c^2$, where $c$ is the speed of ligth.  The related gravitational-wave strain is almost
monochromatic and for a source at $10\, \rm{kpc}$ the maximal
amplitude is $\left. h^{20} \right|_{max} \simeq 2.18 \times 10^{-22}
\sin ^2 \theta$.

\begin{table}
\begin{center}
\caption{\label{tab:modes} Comparison of the first three $l=0,2$
  ordinary and superfluid mode frequencies and
  the~\citet{2002A&A...393..949P} results. The star is the
  non-rotating C0 model, which corresponds to model III
  of~\citet{2002A&A...393..949P}, where the entrainment parameter
  $\bar \veps$ is zero. Frequencies are given in units of $\sigma /
  \sqrt{G \rho_0}$ and have been determined with an FFT of the
  gravitational strain time evolution.  For this specific numerical
  simulation the frequency error bar is $ \Delta \sigma / \sqrt{G
    \rho_0} = 5.58 \times 10^{-3}$.  The Prix and Rieutord values are
  denoted by PR. In the final column we show the relative error
  between our and PR results.}
\begin{tabular}{c c c c   }
\hline
    Mode  &  $ \sigma / \sqrt{G \rho_0} $  &   $ \sigma / \sqrt{G \rho_0} $ \vspace{0.1cm}  & $ \Delta \sigma / \sigma  $ \\
          &      PR                        &                 &  [ \% ]  \\
\hline
   $\rm{F}^\ord$        &  1.91361  &  1.92743  \vspace{0.2mm}  &    0.7 \\
   $\rm{F}^\s$          &  2.52823  &  2.53376  \vspace{0.2mm}  &    0.2 \\
   $\rm{H}_1^{\ord}$    &  3.94917  &  3.94911   \vspace{0.2mm} &   \hspace{-0.3cm}$<$ 0.1\\
   $\rm{H}_1^\s$        &  4.20552  &  4.20420  \vspace{0.2mm}  &   \hspace{-0.3cm}$<$ 0.1\\
   $\rm{H}_2^\ord$      &  5.61069  &  5.52870  \vspace{0.2mm}  &    1.5 \\
   $\rm{H}_2^\s$        &  5.93799  &  5.92165  \vspace{0.2mm}  &   0.3\\  \\

   $\rm{f}^{~\ord}$     &  1.33511  &  1.33178  \vspace{0.1mm}  &    0.2\\
   $\rm{f}^{~\s}$       &  1.83142  &  1.82281  \vspace{0.1mm}  &   0.5 \\
   $\rm{p}_1^\ord$      &  3.47686  &  3.48786  \vspace{0.2mm}  &   0.3 \\
   $\rm{p}_1^\s$        &  3.68465  &  3.69878  \vspace{0.2mm}  &   0.4 \\
   $\rm{p}_2^\ord$      &  5.24187  &  5.25802  \vspace{0.2mm}  &   0.3  \\
   $\rm{p}_2^\s$        &  5.51946  &  5.52876  \vspace{0mm}   &    0.2 \\
\hline
\end{tabular}
\end{center}
\end{table}

With our 2D Newtonian code, we then evolve in time non-radial oscillations
of a superfluid non-rotating star for both the EoS~(\ref{eq:EoSA})
and~(\ref{eq:EosPR}).  The kinetic energy of oscillating superfluid
stars can be determined by the following expression:
\begin{equation}
 E _k = \frac{1}{2} \int d \mtb{r} \left [ \rho_\n \left( 1 -
  \veps_\n \right) | \delta \mtb{v}_{\n} |^2 + 2 \rho_\n \veps_\n
  \delta \mtb{v}_{\n} \cdot \delta \mtb{v}_{\p} + \rho_\p \left( 1 - \veps_\p
  \right) | \delta \mtb{v}_{\p} |^2 \right] \, .
\label{eq:Ek-def}
\end{equation}
If we evolve oscillations that have the same pulsational kinetic
energy as in the case studied with the relativistic code, we obtain $ \left. h^{20}
\right|_{max} \simeq 1.55 \times 10^{-22} \sin ^2 \theta $ for model
A0 and $\left. h^{20} \right|_{max} \simeq 1.433 \times 10^{-22} \sin
^2 \theta $ for model C0. In this calculation, we have used
equation~(\ref{eq:st-unit}) with the parameters of the relativistic
stellar model.
This test shows that we can be confident that the implementation of the
quadruple formula in our code provides reasonable results, in
accordance with the expected  relation between the
pulsational kinetic energy and the gravitational-wave strain.

\section{Results} \label{sec:Res}

Having formulated the time-evolution problem and described our implementation of the
gravitational-wave extraction, we will now discuss our results.
In this section, we focus on the effects  of the
gravitational potential perturbation and the mutual friction force
on axisymmetric and
non-axisymmetric oscillations. We also
provide a more detailed analysis of the
gravitational-wave signal generated by the basic glitch model that we
discussed in a previous work~\citep{2010MNRAS.tmp..554S}.

The pulsation dynamics is studied with a numerical code that evolves
in time the system of hyperbolic perturbation
equations~(\ref{eq:dVxdt})--(\ref{eq:drhoxdt}), solving at each time
step the perturbed Poisson equation~(\ref{eq:dPhi-poiss}). The part of
the code that evolves the hyperbolic equations uses the same
technology as in previous
work~\citep{2009MNRAS.396..951P, 2009MNRAS.394..730P}, whereas the
elliptic equation~(\ref{eq:dPhi-poiss}) is solved using a pseudo
spectral method. The numerical grid
is two-dimensional and covers the volume of the star, i.e. the region
$ 0 \leq r \leq R(\theta) $ and $ 0 \leq \theta < \pi / 2$. The
implementation uses a new radial coordinate $x = x(r,\theta)$, which
is fitted to surfaces of constant chemical potential. This allows us
to consider stars that are highly deformed by rotation.  The
perturbation variables are discretized on this grid and updated in
time with a Mac-Cormack algorithm. The numerical simulations are
stabilised from high frequency noise with the implementation of a
fourth order Kreiss-Oliger numerical dissipation. More technical
details have been discussed in~\cite{2009MNRAS.396..951P, 2009MNRAS.394..730P}.
\begin{figure}
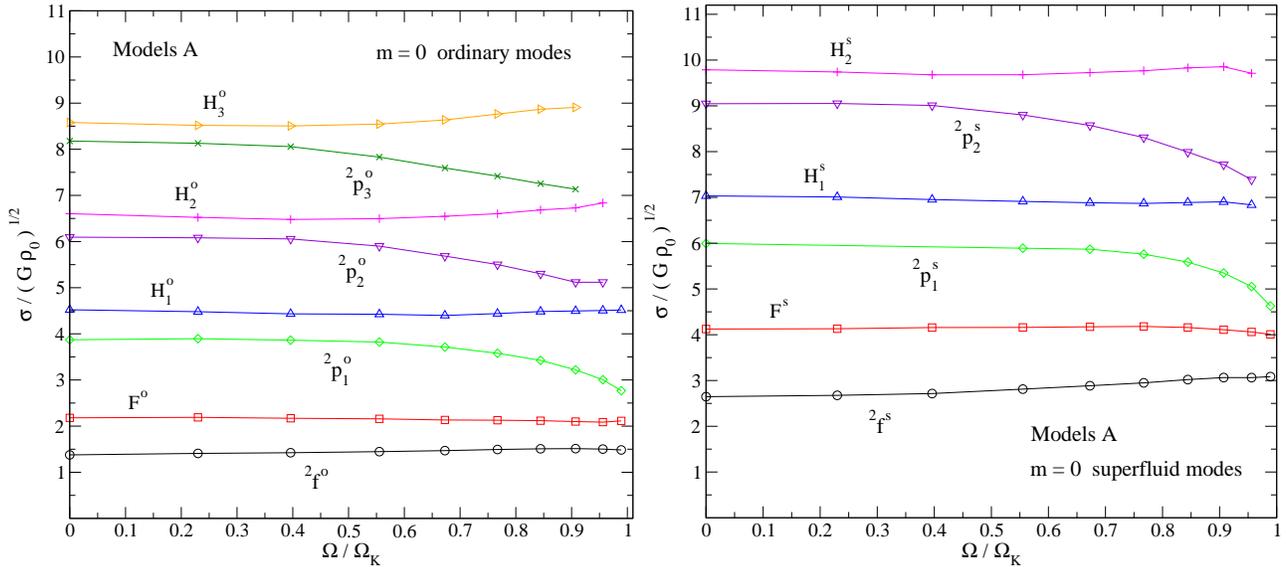

\begin{center}
\includegraphics[height=75mm]{fig5a.eps}
\hspace{0.2mm}\includegraphics[height=75mm]{fig5b.eps}
\caption{This figure displays the effect of rotation on the
quasi-radial and axisymmetric quadrupole modes. We use the sequence of
models A, with entrainment parameter $\bar \varepsilon = 0.5$, proton
fraction $x_\p = 0.1$ and vanishing symmetry energy term. On the
horizontal axis the angular velocity is rescaled with the Kepler
angular velocity $\Omega_K$, while the mode frequencies are given
in dimensionless units and for a rotating frame. In the left panel,
we show some ``ordinary modes'', which are due to the co-moving
degrees of freedom. We identify the $l=0$ and $l=2$ fundamental modes
and the first three quasi-radial overtones and $l=2$ pressure
modes. In the right panel, we show instead the ``superfluid modes'',
which correspond to the counter-moving degrees of freedom. In this
case we show the modes up to the second overtones. For
these non-stratified models, the ordinary and superfluid modes are
decoupled. \label{fig:m0A}}
\end{center}
\end{figure}

In order to solve the elliptic equation~(\ref{eq:dPhi-poiss}) with a
spectral method, and save computational time, we set up a second numerical grid with lower
resolution. This is important, since
the spectral solver must be used at each time step, leading to a
significant slow-down of the simulations. However, the lower
resolution on the spectral grid does not affect the results, as
spectral elliptic solvers provide highly accurate and rapidly
convergent solutions already for relatively coarse
grids~\citep{lrr-2009-1}.  Therefore, at each time step we first fit
the mass density perturbation~$\delta \rho$ on the spectral grid and
then use the spectral routines to determine the gravitational
potential perturbation $\delta \Phi$. Subsequently, we fit the new
value of $\delta \Phi$ to the original grid for the hyperbolic
equations and carry on the evolution.  The numerical code provides
stable simulations for all rotating stellar models considered in this
paper.

In this work, our choice of variables differs from that of~\cite{2009MNRAS.396..951P}. We evolve the velocity perturbations of the two-fluids
components instead of the ``mass flux'' perturbations of the co-moving
and counter-moving degrees of freedom. The two formulations
are obviously mathematically equivalent, but we wanted to develop a code based on the new set
of variables in order to explore which formulation is best suited for
future extensions. This is important, as we plan to add more realistic physics to
our  models by implementing an elastic crust region.  As a first test,
we compare the results of the new code to those obtained in Cowling
approximation by~\cite{2009MNRAS.396..951P}. Neglecting the
perturbation of the gravitational potential, i.e.  setting $\delta \Phi = 0$, we find a
complete agreement between the two numerical codes.

In order to study the spectral properties discussed below, in
Sec.~\ref{sec:spectrum} and~\ref{sec:MFspec}, we consider ``generic'' initial
conditions that excite a large set of oscillation modes. For Type I
perturbations we provide the following expression for the mass
density:
\begin{equation}
\delta \rho_{\n} = - \delta \rho_{\p} = \left( \frac{r}{ R\left( \theta
           \right) }\right)^l Y_{l l} \left(\theta,\phi\right) \, ,
           \label{eq:fIP}
\end{equation}
where neutrons and protons are initially  counter-moving.  For
Type II perturbations, we excite mainly normal and superfluid
r-modes with the following initial data:
\begin{equation}
\delta \mtb{v}_{\n} = - \delta \mtb{v}_{\p} = \left( \frac{r}{ R\left( \theta
           \right) }\right)^l Y_{l l}^B \left(\theta,\phi\right) \, ,
\end{equation}
where $Y_{l l}^B \left(\theta,\phi\right)$ is a magnetic spherical
harmonic~\citep{1980RvMP...52..299T}.  For the glitch simulations, we
use the non-corotating solutions derived in
Sec.~{\ref{sec:Noncorback}.

We test the elliptic solvers by comparing the mode frequencies
extracted from our time evolutions to those obtained in the
frequency domain by~\cite{2002A&A...393..949P}.  We determine the
oscillation frequencies of the non-rotating model C0~(see
Table~\ref{tab:back-models}), which corresponds to model III
of~\cite{2002A&A...393..949P}. For the zero entrainment case,
i.e. when $(\bar \veps = 0)$, the results in Table~\ref{tab:modes} show that the
frequencies determined with our code (by an FFT of the time-evolved
perturbations) agree very well with those calculated
by~\cite{2002A&A...393..949P}.

\begin{figure}
\begin{center}
\includegraphics[height=75mm]{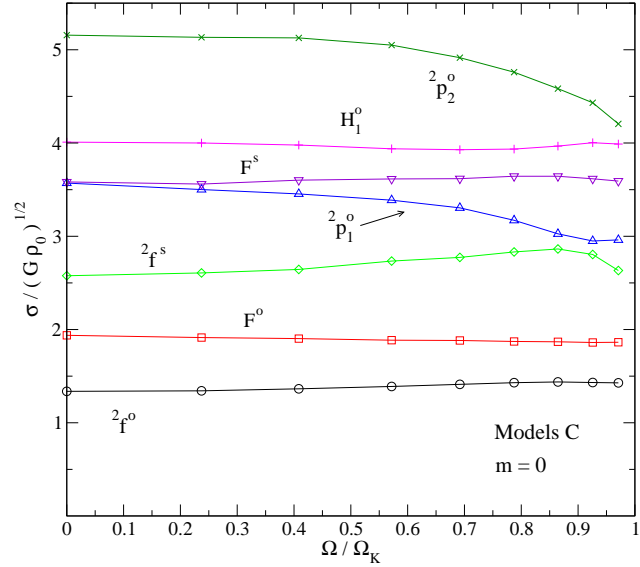}
\caption{ In this figure, we show the axisymmetric modes for the
  sequence of C models determined in a rotating frame. These stellar
  models are stratified and the ordinary and superfluid degrees of
  freedom are coupled. We identify some of the acoustic modes and
  their dependence on the rotational rate $\Omega/\Omega_K$. \label{fig:axyC}}
\end{center}
\end{figure}

\subsection{Spectrum} \label{sec:spectrum}

The oscillation spectrum of superfluid rotating neutron stars contains
the imprints of two-fluid dynamics and of the mutual friction force.
For a single fluid star, the general mode classification is based on
the main restoring force that acts on the displaced fluid
elements~\citep{Cowling:1941co}. For nonrotating models without
magnetic field and crust, the spectrum is formed by the acoustic, the
fundamental and the gravity modes. The acoustic modes are mainly
restored by pressure variations and cover the high frequency range of
the spectrum, above 1~kHz. At lower frequencies, typically below
100~Hz, composition and thermal gradients generate the class of
gravity modes that are restored by buoyancy.  The fundamental mode,
whose frequency scales with the average stellar density, separates
these two classes of modes.  In rotating stars, the Coriolis force
provides an additional restoring force, leading to the presence of
inertial modes. Since the frequency of these modes scales with the
rotation rate, they typically lie in the same low frequency region as
the g-modes. For rotating stars with composition gradients, the
inertial and gravity modes form a unique class with mixed properties,
referred to as gravity-inertial modes~\citep[for a recent analysis
see][]{2009MNRAS.394..730P, 2009PhRvD..80f4026G}.

In addition to this general classification, any oscillation mode can
be labeled by the indices $(l,m)$ associated with the spherical
harmonics $Y_l^m(\theta,\phi)$. In spherical stars, this is due to the
decomposition of the perturbation functions in vector harmonics.  For
rotating stars, we can use the same description as long as we can
track a mode back to its non-rotating limit.  Finally, for any value
of $(l,m)$, the oscillation modes can be ordered by the number of
radial nodes in their eigenfunctions.  The fundamental mode ${}^l
\textrm{f}$ does not have radial nodes, while the series of pressure
${}^l\textrm{p}_i$ and gravity modes ${}^l\textrm{g}_i$ have $i$
nodes.

In superfluid neutron stars, the additional degree of freedom enriches
the dynamics. The two fluids can oscillate both in phase and
counter-phase. The co-moving degree of freedom produces the class of
``ordinary modes'', very similar to the single fluid results described
above. There is, however, one important difference: the gravity modes
are absent in superfluid stars~\citep{1995A&A...303..515L,
2001MNRAS.328.1129A, 2002A&A...393..949P}. The counter-moving degree
of freedom generates a new class of acoustic and inertial modes, known
as ``superfluid'' modes. These modes strongly depend on the superfluid
aspects, such as entrainment and mutual friction. We will label
ordinary and superfluid modes by an upper index, for instance the
$l=2$ fundamental ordinary mode will be expressed as ${}^2
\textrm{f}^{\hspace{0.6 mm} \rm{o}}$, while ${}^2
\textrm{f}^{\hspace{0.6 mm} \rm{s}}$ represents the corresponding
superfluid mode.

We focus our attention on the quasi-radial ($l=0$) and quadrupole
($l=2$) oscillation modes and study their behaviour in rapidly
rotating models all the way to the mass shedding limit.  In
non-rotating models, the $l=0$ modes are purely radial and do not
generate gravitational radiation. However, due to coupling of the
different multipoles, this is no longer true in the rotating case.
The quasi-radial fundamental mode will be denoted by F and its $i$
overtones by $\textrm{H}_i$.  The quadrupole modes ($l=2$) are
expected to be dominant in the gravitational signal, and we study both
axisymmetric and non-axisymmetric oscillations. These correspond to
$m=0$ and $m=2$ respectively.

We start by considering the axisymmetric oscillations for the two
sequences of rotating models A and C.  For a small velocity lag
between the two fluids, the entrainment parameter $\bar \veps$ can be
chosen independently from the background model (see
Section~\ref{sec:EoS}). Recent work suggests that it can assume values
in the range $0.2 \leq \bar{\veps} \leq
0.8$~\citep{2008MNRAS.388..737C}. Here, we consider only the case
$\bar \veps = 0.5$, as the effect of this parameter on the oscillation
frequencies has been already discussed
elsewhere~\citep{2002A&A...393..949P, 2009MNRAS.396..951P,
2009MNRAS.397.1464H}. The parameters~$\veps_\x$ for the two fluids are
then given by $\veps_\n = x_\p \, \bar \veps$ and $\veps_\p = \bar
\veps - \veps_\n$. For models A we must also specify the proton
fraction and the symmetry energy term. These are, respectively, set to
$x_\p=0.1$ and $\sigma = 0$. For a discussion of the effect of
$\sigma$ on the spectrum, see~\citet{2009MNRAS.396..951P}. From the
numerical simulations we determine the mode frequencies with an FFT of
the time-evolved perturbation variables. In order to identify the
different modes, we use also the eigenfunction extraction technique
developed by~\cite{Stergioulas:2003ep} and~\cite{Dimmelmeier:2005zk}.

In Fig.~\ref{fig:m0A} we show, for the non-stratified A models, some
of the axisymmetric frequencies of the quasi-radial ($l=0$) and
quadrupole ($l=2$) modes. In the left and right panels we show the
``ordinary'' and ``superfluid'' modes, respectively. These two mode
families are decoupled in non-stratified stars, and in fact the results
in Fig.~\ref{fig:m0A} do not hint at any interaction in the spectrum. However,
within the sets of ordinary and superfluid modes avoiding
crossings may appear.  For instance, the ordinary quasi-radial mode
$\rm{H}_1^\ord$ and the ordinary pressure mode
${}^{2}\rm{p}_{2}^{\ord}$ seem to have an avoiding crossing when the
star is rotating at $90\%$ of the mass shedding limit.
The effects of the chemical coupling on the spectrum is evident in
Fig.~\ref{fig:axyC}, where we show some of the axisymmetric modes for
the C models. In this case, the superfluid fundamental mode
${}^{2}\rm{f}^{\hspace{0.6mm}\s}$ and the ordinary first pressure mode
${}^{2}\rm{p}_{1}^{\ord}$ interact through an avoiding crossing near 90\%
of the Kepler limit.

For a given $l$ multipole, the non-axisymmetric modes of
non-rotating stars have a degeneracy with respect to $m$. Rotation
removes this degeneracy and splits each mode into $2l+1$ distinct branches. Besides
the $m=0$ case considered above, we consider the $|m|=2$ modes that
have a pro- and retro-grade motion with respect to the star.  In
Fig.~\ref{fig:Nonaxy}, we show the frequencies of the acoustic modes
for models A (left panel) and C (right
panel). The present results improve on the analysis of~\citet{2009MNRAS.396..951P}, that studied the
dependence of the rotational splitting on the entrainment parameter
within the Cowling approximation. The main improvement concerns the
introduction of the gravitational potential perturbations. However, this does not
alter the qualitative
effects of rotation on
the splitting.

\begin{figure}
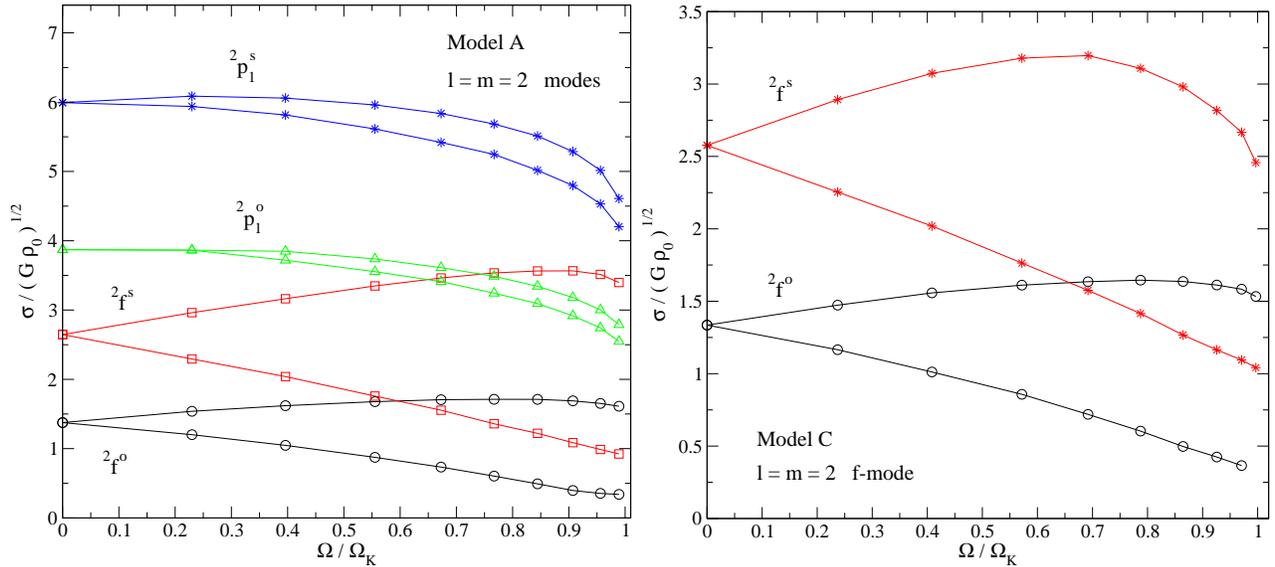

\begin{center}
\includegraphics[height=75mm]{fig7a.eps}
\hspace{0.2mm}\includegraphics[height=75mm]{fig7b.eps}
\caption{ In this figure, we show the rotational splitting of the
$l=m=2$ non-axisymmetric modes as measured in the rotating frame. The
left and right panels display the modes for the sequence of A and C
models respectively.  \label{fig:Nonaxy} }
\end{center}
\end{figure}

\subsection{Mutual friction effects on the spectrum} \label{sec:MFspec}

In order to study the effects of the mutual friction force on the
oscillation spectrum, it is useful to write the momentum equation for
the relative motion between protons and neutrons. We can combine the
Euler-type equations~(\ref{eq:dVxdt}) and obtain the following
expression in the rotating frame:
\begin{equation}
\left( 1 - \bar \veps \right) \partial_{t} \delta \mtb{w}^{\p\n} = -
\nabla \left( \delta \tilde \mu_{\p} - \delta \tilde \mu_{\n} \right)
- 2 \mathcal{\bar{B}}' \mtb{\Omega} \times \delta \mtb{w}^{\p\n} + 2
\mathcal{\bar{B}} \, \hat{\mtb{\Omega}} \times \mtb{\Omega} \times
\delta \mtb{w}^{\p\n} \ , \label{eq:w}
\end{equation}
where we have defined
\begin{equation}
 \mathcal{\bar{B}}' \equiv 1 -\frac{ \mathcal{B}'}{x_\p}   \, , \qquad \qquad \mathcal{\bar{B}} \equiv \frac{ \mathcal{B}}{x_\p} \, .
\end{equation}
Equation~(\ref{eq:w}) makes the effects of the
mutual friction parameters $\mathcal{B}$ and $\mathcal{B}'$  more evident. The term that includes $\mathcal{B}$ is
dissipative and tends to damp the relative motion, and consequently mainly affects
the superfluid modes. If the co- and counter-moving degrees of freedom
are coupled, for instance due to the EoS, the mutual
friction dissipation affects also the ordinary modes. Results to this effect, have been provided by~\citep{1995ApJ...444..804L, 2009PhRvD..79j3009A} for the f-modes and~\citep{2000PhRvD..61j4003L,2003ApJ...586..403L,
2009MNRAS.397.1464H} for the r-modes.  The term proportional
to $\mathcal{B}'$ modifies the Coriolis force, as one can
see in equation~(\ref{eq:w}). Its effects are not dissipative,
but may change the frequencies of the superfluid modes.  This is certainly
expected in the case of the inertial modes as they are rotationally
restored, but we will see that  the non-axisymmetric fundamental
modes can also be affected.

The magnitude of the mutual friction can be studied by introducing a
dimensionless drag parameter $\mathcal{R}$, defined by~\citep{2009MNRAS.397.1464H}:
\begin{equation}
\mathcal{B} = \frac{\mathcal{R}}{1+\mathcal{R}^2} \, , \qquad \qquad
\mathcal{B}' = \frac{\mathcal{R}^2}{1+\mathcal{R}^2} \, .
\label{eq:Rdef}
\end{equation}
Two extreme drag regimes can then be discerned for the mutual friction
force. In the ``weak'' drag regime $\mathcal{R} \ll 1$, whereas
the ``strong'' drag regime corresponds to $\mathcal{R} \gg 1$.
The most commonly considered cause of mutual friction is the scattering of the electrons
off the magnetic field of the neutron vortices. This mechanism is firmly in the weak drag
regime, where~$\mathcal{B}\ll 1$ and $\mathcal{B}' \ll
\mathcal{B}$. In this case, we expect the mutual friction to act mainly on
the mode damping. It should have negligible effects on the oscillation frequencies themselves.
%
\begin{figure}
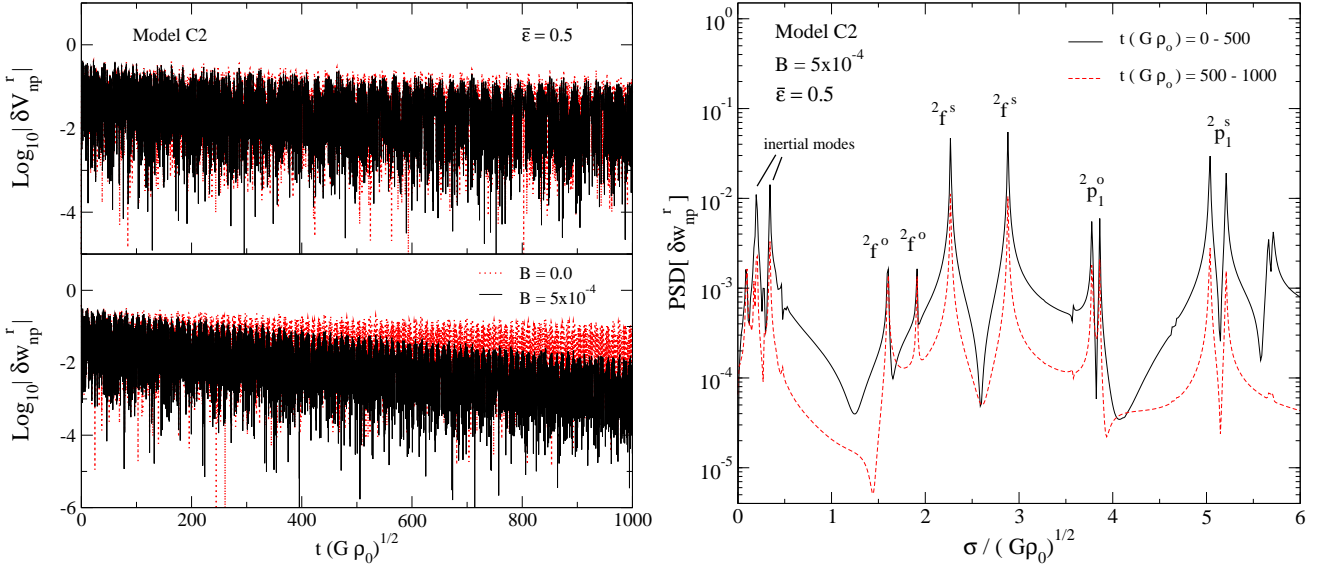

\begin{center}
\includegraphics[height=75mm]{fig8a.eps}
\hspace{0.2mm}\includegraphics[height=75mm]{fig8b.eps}
\caption{This figure illustrates the effect, in the weak drag regime,
  of the mutual friction on the stellar oscillations.  For the model
  C2 with $\bar{\veps} = 0.5$, the left panel displays the radial
  component of the variables $\delta \mtb{V} _{\n\p}$ (upper panel)
  and $\delta \mtb{w}_{\n\p}$ (lower panel) for two long simulations
  with $\mathcal{B}=0$ and $\mathcal{B}=5\times10^{-4}$,
  respectively. The horizontal axis shows the dimensionless
  evolution time.  The lower-left panel shows that the counter-moving
  degrees of freedom are damped due to mutual friction dissipation. In
  the right panel, we show an FFT of the function $\delta
  w_{\n\p}^{r}$ for the $\mathcal{B}=5\times10^{-4}$ case. In order to
  study the mode amplitude variation with time, we have performed an
  FFT of the first part of the simulation, where $0 \leq t \left( G
  \rho_{0} \right) \leq 500$ (solid line), and subsequently of the
  second part, where $ 500 \leq t \left( G \rho_{0} \right) \leq 1000$
  (dashed line). On the horizontal axis is shown the dimensionless
  mode frequency $\sigma / \left( G \rho_0 \right)^{1/2}$, as measured
  in the rotating frame.  As expected, the superfluid modes exhibit a
  faster damping than the ordinary modes.
\label{eq:weak}}
\end{center}
\end{figure}

Recent discussions suggest that the strong drag regime may lead to interesting,
potentially important, results~\citep{2009MNRAS.397.1464H, 2009PhRvD..79j3009A}.
Since our level of theoretical understanding is not sufficient to rule out this
case, we also consider the $\mathcal{R}\gg 1$ regime.  From
equations~(\ref{eq:Rdef}) we see that in the strong drag regime
$\mathcal{B}' \simeq 1$ and $\mathcal{B} \ll \mathcal{B}'$. The main
effect should then be on the mode frequencies, while the
dissipation can be considered negligible. In principle, we could explore
also the intermediate regime, where $\mathcal{R} \simeq 1$ and both
energy dissipation and frequency changes are
important. However, this case is essentially a combination of the effects
that we can study in the weak and strong regimes. Hence, we do not consider
the  intermediate regime in this work.

\subsubsection{Weak drag regime}

Let us first consider the weak drag regime by evolving in time the
oscillations of model C2. In Fig.~\ref{eq:weak} we show the results
from two long simulations where we have fixed $\mathcal{R} = 0$ and
$\mathcal{R} = 5\times10^{-4}$, respectively. In the left panel, we
show the grid-averaged value of the velocities $\delta \mtb{V}_{\n\p}
= \delta \mtb{v}_{\n} + \delta \mtb{v}_{\p}$ and $\delta
\mtb{w}_{\n\p}$. In the upper-left panel, the two curves appear
similar showing a weak damping that is mainly due to the numerical
dissipation. In fact, the quantity $\delta \mtb{V}_{\n\p}$ describes
the evolution of the co-moving degree of freedom, which is weakly
affected by the weak mutual friction. Looking
more carefully at the results, we note that some damping is present
in the $\mathcal{R} = 5\times10^{-4}$ case. This is due to the
chemical coupling with the counter-moving degree of freedom, which
is strongly damped. This is evident from the results in the lower-left panel
of Fig.~\ref{eq:weak}, which show that the amplitude of
the relative velocity $\delta \mtb{w}_{\n\p}$ decreases during the
evolution.

These results suggest that, as expected, superfluid modes are damped
faster than the ordinary modes.  In order to study how the mode
amplitude changes during the evolution, we divide the time-evolved
data into two equal sets and perform an FFT for each part.
Results for the variable $\delta \mtb{w}_{\n\p}$ in the $\mathcal{R} =
5\times10^{-4}$ case are shown in the right panel of
Fig.~\ref{eq:weak}. We see that the superfluid fundamental and first
pressure modes are damped faster then their ordinary counterparts.

The effect of the mutual friction has also been tested
by~\cite{2010MNRAS.tmp..554S}, by comparing the glitch spin-up
time extracted by our numerical evolutions against an analytical
formula derived within a body-averaged approximation.

While our results demonstrate good progress, they are not quite satisfactory
in one important respect. Ideally, one would like to be able
to extract both oscillation frequency and damping time for the different modes
seen in the evolution. However, so far
we have not managed to extract the mutual friction damping rate of individual
oscillation modes with the desired precision. This is basically because of the
fact that the damping is very slow. It is also sensitive
  to the velocity lag between the two fluid components. At the present
  time it is not clear to us whether a time-evolution code provides
   a useful alternative to frequency-domain calculation for the
   damping-rate problem.

\begin{figure}
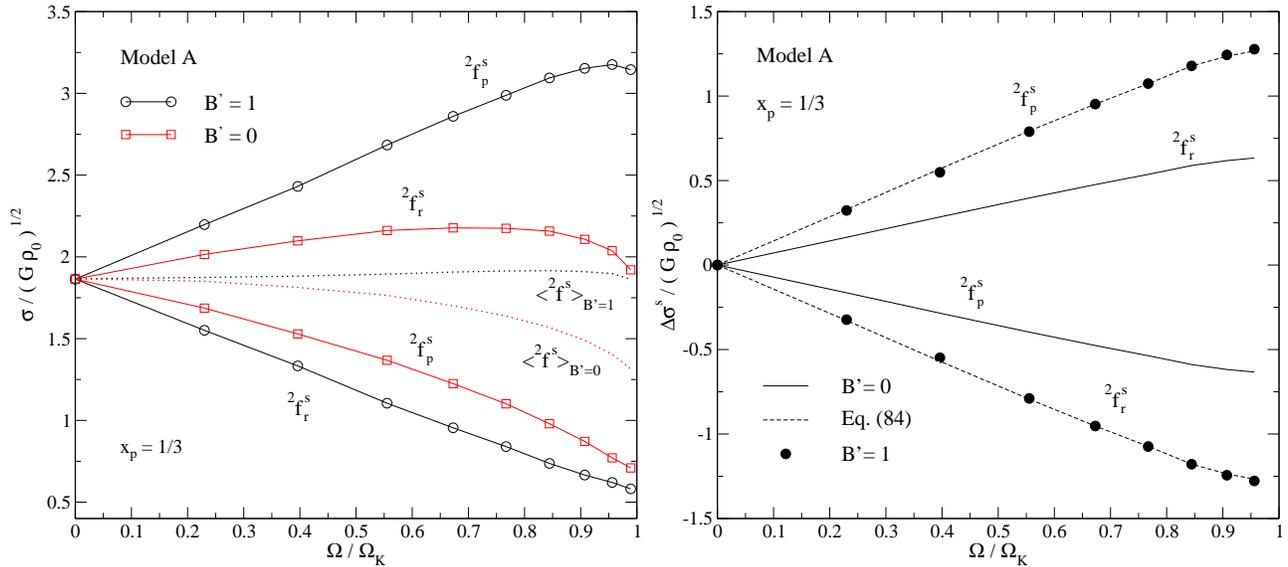

\begin{center}
\includegraphics[height=75mm]{fig9a.eps}
\hspace{0.2mm}\includegraphics[height=75mm]{fig9b.eps}
\caption{ This figure shows the effects of mutual friction, in
  the strong drag regime, on the rotational splitting of the
  superfluid $l=m=2$ f-mode. The axis labels are shown in
  dimensionless units and the mode frequencies are determined in the
  rotating frame. For the sequence of models A with $\bar \veps = 0$
  and $x_{\p} = 1/3$, we study the $\mathcal{B}'=0$ and
  $\mathcal{B}'=1$ cases. In the left panel, we show the effects of
  mutual friction on the pro-grade ${}^2\rm{f}^{\s}_{p}$ and
  retro-grade ${}^2\rm{f}_{r}^{\s}$ modes, respectively, and the
  averaged frequencies $\langle \sigma \rangle = \left( \sigma_{p} +
  \sigma_{r} \right) / 2 $ of the two mode patterns. In the right
  panel, we show the deviation of the $l=m=2$ $\rm{f}^{s}$ mode
  defined by equation~(\ref{eq:dev1}). The dashed lines are
  determined by using equation~(\ref{eq:dev2}). These empirical relations agree very
  well with the values of the frequency deviation for the
  $\mathcal{B}'=1$ case, which are shown with filled circles.
 \label{fig:f2-R} }
\end{center}
\end{figure}
\subsubsection{Strong drag regime}

Next we explore the effects of the mutual friction in the strong drag regime, focussing on
the $l=m=2$ superfluid f- and r-modes. The parameter $\mathcal{B}'$
now dominates  the mutual friction force affecting the Coriolis
term in equation~(\ref{eq:w}).

The first aspect we want to understand is whether the rotational
splitting of the superfluid f-mode is modified by the $\mathcal{\bar
B}'$ parameter. Based on our expectations,  we assume that the frequency of
the ${}^2\rm{f}^{\s}$ mode is described by the following relation up to
order $\Omega^2$;
\begin{equation}
\sigma^{\s} = \sigma_{NR}^{s} + c_{1} \left( \bar \veps , \sigma ,
\mathcal{\bar{B}}', m \right) \Omega + \mathcal{O} \left( \Omega^2
\right) \, ,  \label{eq:split0}
\end{equation}
where $c_{1}$  depends on the azimuthal index $m$
and the stellar parameters $\bar{\veps}, \sigma$ and
$\mathcal{\bar{B}}'$. For $\mathcal{\bar{B}}'=0$, we have already
studied the dependence of the ${}^2\rm{f}^{\s}$ mode on the
entrainment parameter $\bar \veps$ and the symmetry energy term
$\sigma$~\citep{2009MNRAS.396..951P}. Therefore, we focus on the
$\bar \veps=\sigma=0$ case and vary the parameter
$\mathcal{\bar{B}}'$. Using our previous results, we can
re-write~equation~(\ref{eq:split0}) as follows:
\begin{equation}
\sigma^{\s} = \sigma_{NR}^{s} + \mathcal{\bar{B}}' \Omega +
\mathcal{O} \left( \Omega^2 \right) \, , \label{eq:sig-split}
\end{equation}
We can then test this result against the numerical simulations.

\begin{figure}
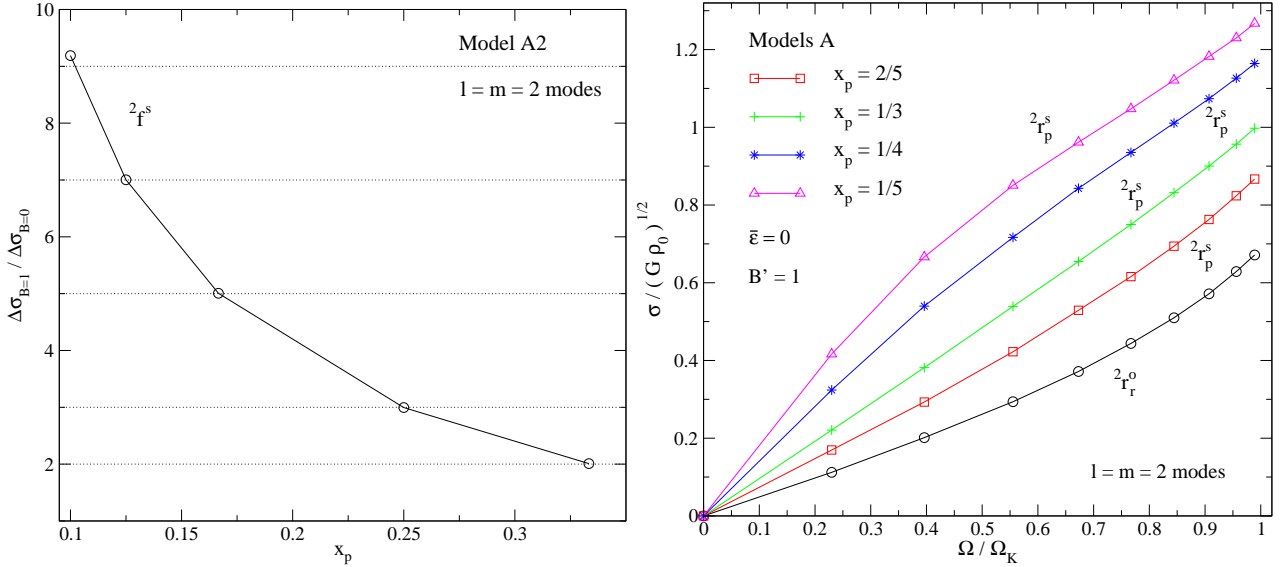

\begin{center}
\includegraphics[height=75mm]{fig10a.eps}
\hspace{0.2mm}\includegraphics[height=75mm]{fig10b.eps}
\caption{ For model A2, we show (in the left panel) how the $l=m=2$
  superfluid f-mode depends on the proton fraction in the strong drag
  regime with $\mathcal{B}'=1$.  The vertical axis displays the ratio
  of the quantity $\Delta \sigma$ for the $\mathcal{B}'=1$ and
  $\mathcal{B}'=0$ cases, see equation~(\ref{eq:dev2}). In the right
  panel, we show the $l=m=2$ ordinary ${}^2\rm{r}^{\ord}$ and
  superfluid ${}^2\rm{r}^{\s}$ modes for the sequence of models A with
  $\veps=\sigma=0$ and $\mathcal{R}=10^{3}$ ($\mathcal{B}'=1$). In the
  strong drag regime, the superfluid ${}^2\rm{r}^{\s}$ mode exhibits a
  strong dependence on the proton fraction $x_\p$ and thus on the
  parameter $\mathcal{\bar{B}}'$.
\label{fig:f2-Bp-scal}}
\end{center}
\end{figure}

We first study a sequence of A models with $x_{\p}= 1/3$,
$\mathcal{R}=10^{3}$, and  $\mathcal{\bar{B}}' = -2$. According
to equation~(\ref{eq:sig-split}), we would expect  the pro- and
retro-grade mode-branches to be exchanged compared to the
$\mathcal{R}=0$ case.  In Fig.~\ref{fig:f2-R} we show the
${}^2\rm{f}^{\s}$ mode for stellar models A  rotating up to the mass
shedding limit with $\mathcal{R}=0$ and $\mathcal{R}=10^{3}$,
respectively.  The results show that (\ref{eq:sig-split})  describes  the ${}^2
\rm{f}^{\s}$ mode very well in the strong drag regime. We find that
the scaling is quite accurate for stars up to $\Omega/\Omega_{K} = 0.9$ (note that
this analysis is not reported in Fig.~\ref{fig:f2-R}). However, the agreement is
not so good when
the mutual friction vanishes. It seems
that for $\mathcal{R}=0$ the effects of the centrifugal force
becomes important for slower rotating models than in the $\mathcal{R}=10^{3}$
case. This behaviour is evident in Fig~\ref{fig:f2-R}, when we consider
the averaged frequency between the $m=2$ pro- and retro-grade modes,
i.e. $\langle \sigma \rangle = \left( \sigma_{\rm{p}} +
\sigma_{\rm{r}} \right) / 2$.

However, we can take into account the effects of the centrifugal force
on the average mode frequency and determine a connection between the
superfluid f-mode frequencies in the strong and weak drag regimes. To
this end, we define the mode deviation from its averaged value:
\begin{equation}
\Delta \sigma^{\s} = \sigma^{\s} - \langle \sigma^{\s} \rangle   \, ,
\label{eq:dev1}
\end{equation}
We then expect, from equation~(\ref{eq:sig-split}), that the following
relation is valid:
\begin{equation}
\Delta \sigma^{\s}_{\mathcal{B}'= 1} \simeq  \mathcal{\bar{B}'} \Delta \sigma^{\s}_{\mathcal{B}'= 0} \, .
\label{eq:dev2}
\end{equation}
In the left panel of Fig.~\ref{fig:f2-R}, we show the quantity $\Delta
\sigma$ for the ${}^2\rm{f}^{\s}$ mode in the strong drag regime and
for vanishing mutual friction. The results for the $\mathcal{R}=10^3$ case
agree very well with the values obtained from
equation~(\ref{eq:dev2}).

So far, we  have studied a sequence of rotating stars with fixed proton
fraction. Now, we test relation~(\ref{eq:dev2}) by varying $x_\p$ and
choosing the rotational model A2 with
$\mathcal{R}=10^3$. The results in Fig.~\ref{fig:f2-Bp-scal} show that the scaling of
the ${}^2\rm{f}^{\s}$ mode with the proton fraction is well described
by equation~(\ref{eq:dev2}). However, when $x_\p=0.1$, there is a small
difference between the numerical and analytical values. This effect
might be due to the second order terms that we have neglected in the
expansion~(\ref{eq:sig-split}). It is natural that these become important when the
parameter $\mathcal{\bar B}'$ is close to 10.

Let us now study the behaviour of the superfluid $l=m=2$ r-mode in the
strong drag regime. Oscillations restored by the Coriolis force generate
the class of inertial modes, which can be classified (by their parity)
as axial-led or polar-led~\citep{1999ApJ...521..764L}. The ordinary r-modes form a sub-set
that is purely axial in the slow-rotation
limit. The superfluid problem is somewhat different in that
a purely axial superfluid r-mode exists only
in non-stratified stars. When composition gradients are present, the
superfluid r-mode acquires a polar component and assumes the nature of a general
inertial mode~\citep{2009MNRAS.397.1464H}.

For a constant density stellar model with $\mathcal{B}=0$, the
frequency of the ${}^2\rm{r}^{\s}$ mode in the rotating frame is described  by the following
relation~\citep{2009MNRAS.397.1464H}:
\begin{equation}
\sigma^{\s} = \gamma_{\veps} \mathcal{\bar{B}}' \sigma^{\ord}
= \frac{2 m \gamma_{\veps} \mathcal{\bar{B}}' \Omega }{l\left(l+1\right) }
\label{eq:r0}
\, ,
\end{equation}
where $\sigma^{\ord}$ is the frequency of the ordinary r-mode. In the case of
$l=m=2$ we have $\sigma^{\ord}_{\rm{r}}= 2 \Omega / 3$. For
compressible models, equation~(\ref{eq:r0})  approximately describes
the frequency of the ${}^2\rm{r}^{\s}$ mode only for slowly rotating
stars. In fact, when a star rotates rapidly the effects of
$\mathcal{O}\left( \Omega ^{3} \right)$ must be taken into
account. In our time-evolutions, the rotational deformation of the
star is completely described by the axisymmetric background. Meanwhile,
 in the slow-rotation approximation the equilibrium
configuration remains spherical and the rotational effects on the
spectrum are described by a perturbation expansion in $\Omega$. By
using a slow-rotation approximation up to $\mathcal{O}\left( \Omega
^{3} \right)$  and the Cowling approximation,
\citet{2009MNRAS.397.1464H} determined the frequency of the
${}^2\rm{r}^{\ord}$ and ${}^2\rm{r}^{\s}$ modes in closed form.  For
the sequence of non-stratified A models with zero mutual friction, we
have compared the r-mode frequencies of~\citet{2009MNRAS.397.1464H} with
the spectrum extracted by the time-evolutions and found an agreement
to better than $3\%$ up to models with $\Omega/\Omega_{K} \simeq
0.77$~\citep{2009MNRAS.396..951P}. For faster rotation, the
slow-rotation approximation would require the calculation of terms of
higher order than $\mathcal{O}\left( \Omega ^{3} \right)$, which can
be computationally prohibitive. In the strong drag regime, the
effects of the higher order pertubative terms can become important
even for relatively slowly rotating models, as a large value of
$\mathcal{\bar{B}}'$ increases the effective strength of the Coriolis force.

We study the superfluid r-modes of the rotating A models, where we fix
the values of the entrainment and symmetry energy to zero, $\bar \veps
= \sigma = 0$. The effects of these two parameters on the r-mode
spectrum have already been  studied by~\citet{2009MNRAS.396..951P}. In
this paper, we focus on the effects of the mutual friction parameter
$\mathcal{\bar{B}}'$ by choosing $\mathcal{R}=10^{3}$ and consider
four values of the proton fraction, namely $x_\p =
2/5,\ 1/3,\ 1/4,\ 1/5$. The extraction of the r-mode frequencies from
the time-evolutions requires longer simulations, as these modes are in
the low-frequency regime.  In order to save computational time, we
adopt the Cowling approximation. In the right panel of
Fig.~\ref{fig:f2-Bp-scal} we show the ordinary ${}^2\rm{r}^{\ord}$ and
superfluid ${}^2\rm{r}^{\s}$ modes for different proton fractions. The
${}^2\rm{r}^{\ord}$ mode has a retro-grade motion with respect to the
star and is not affected by the parameter $x_\p$. In contrast, the
${}^2\rm{r}^{\s}$ mode depends strongly on $x_\p$ and has a pro-grade
nature, as $\mathcal{\bar{B}}'$ is negative.

\begin{table}
\begin{center}
\caption{\label{tab:fit} This table provides the fitting coefficients $c_3, c_5$
   of equation~(\ref{eq:fit}), and their errors $\Delta c_3$ and
   $\Delta c_5$, for the superfluid ${}^2\rm{r}^{\s}$ modes in
   the strong drag regime. The results correspond to the r-modes of four
   sequences of models A with $\mathcal{R}=10^{3}$ and
   $\veps=\sigma=0$, and where the proton fraction takes the values
   shown in the first column. In the second column, we give the value
   of the parameter $\mathcal{\bar{B}}'$. For the first two models, we
   do not show the values of $c_5$, as we fit the r-mode frequencies
   with $c_5=0$.}
\begin{tabular}{c c c c c c }
\hline
  $ x_{\p} $  &  $ \mathcal{\bar{B}}' $  & $c_3$  & $ \Delta c_3 $  & $c_5 $  & $ \Delta c_5 $ \\
             &                          &        &   $ \times 10^{-3}$   & $ \times 10^{-3}$ & $         \times 10^{-3}$   \\
\hline
   2/5          &  -1.5              &    ~0.11338  &     2.266   &  &                            \\
   1/3          &  -2.0              &    ~0.01121  &     0.958   &  &                            \\
   1/4          &  -3.0              &    -0.05599  &     0.948   &  6.275  &  0.233        \\
   1/5          &  -4.0              &    -0.05982  &     1.742   &  4.013  &  0.241      \\
\hline
\end{tabular}
\end{center}
\end{table}

In order to understand the behaviour of the superfluid r-modes, we
assume that for $\bar \veps = \sigma = 0$ the frequency of a
counter-moving $l=m=2$ r-mode is described by the following relation;
\begin{equation}
\frac{\sigma^{\s}} {\sqrt{G \rho_{0} }} = \frac{2}{3}
\mathcal{\bar{B}'} \frac{ \Omega }{\sqrt{G \rho_{0} }} + c_{3} \left(
\mathcal{\bar{B}'} \frac{\Omega }{\sqrt{G \rho_{0} }} \right)^3 +
c_{5} \left( \mathcal{\bar{B}'} \frac{\Omega }{\sqrt{G \rho_{0} }}
\right)^5 + \mathcal{O} \left( \Omega \right) ^7 \, , \label{eq:fit}
\end{equation}
where the frequencies and angular velocities are expressed in
dimensionless units, while $c_3$ and $c_5$ are two fitting parameters.
The numerical spectrum, shown in Fig.~\ref{fig:f2-Bp-scal}, is well
described by the first term of equation~(\ref{eq:fit}) only for
slowest rotating models. For stars with $x_\p > 1/4$ the agreement is
good up to $\Omega/\Omega_{K} \simeq 0.4$, while for $x_\p \leq 1/4$
the range reduces to $\Omega/\Omega_{K} \leq 0.2$. In particular, we
learn from the results in Fig.~\ref{fig:f2-Bp-scal} that the mode
pattern changes concavity for increasing values of
$\mathcal{\bar{B}}'$. This can be an effect of the $\mathcal{O}
\left(\Omega^3 \right)$ and $\mathcal{O} \left(\Omega^5 \right)$ terms
of equation~(\ref{eq:fit}).  Therefore, we fit our numerical data with
equation~(\ref{eq:fit}) and determine the parameters $c_3$ and $c_5$.
For models with $x_\p > 1/4$, a good fit can be determined by setting
$c_5=0$ and calculating only the coefficient $c_3$.  For the ordinary
${}^2\rm{r}^{\ord}$ mode we obtain $c_3 = 0.4852 \pm 0.0126$, while
for the superfluid ${}^2\rm{r}^{\s}$ modes the results are given in
Table~\ref{tab:fit}.  When the proton fraction is smaller, i.e. $x_\p
\leq 1/4$, we must use the entire equation~(\ref{eq:fit}).  The
results of the corresponding fits are given in Table~\ref{tab:fit}. In
particular, when $x_\p \leq 1/4$ we note a sign change in the
parameter $c_3$ that represents the concavity variation of the mode
pattern.

\begin{figure}
\begin{center}
\includegraphics[height=75mm]{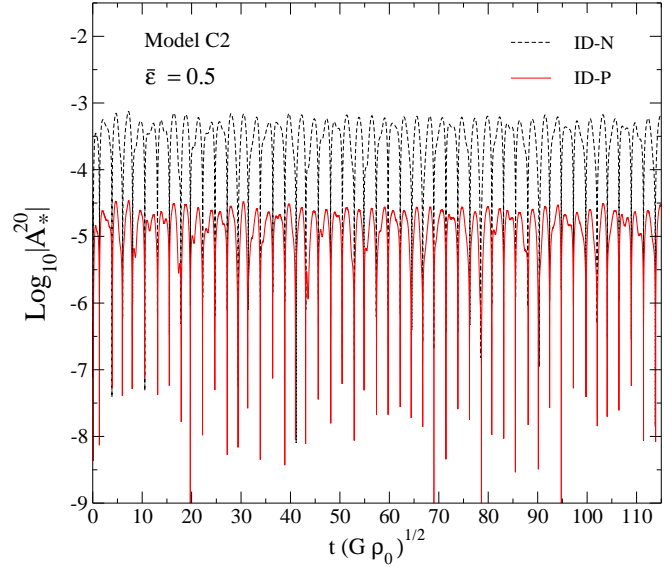}
\caption{This figure shows the waveform of the gravitational-wave
  signal for the C2 model with constant entrainment parameter $\bar
  \veps = 0.5$. We show the evolution of the two independent initial
  conditions ID-N (dashed-line) and ID-P (solid-line). The axis labels
  are in dimensionless units.
\label{fig:A20}}
\end{center}
\end{figure}

\subsection{Glitch gravitational signal} \label{sec:GGW}

We now turn to the gravitational-wave signal generated by
 initial axisymmetric configurations such that the protons and the
neutrons  rotate with a velocity lag. As discussed in
Section~\ref{sec:Noncorback}, these
configurations can be determined with the perturbative approach
developed by~\cite{2004MNRAS.347..575Y}.  We have already
considered this problem~\citep{2010MNRAS.tmp..554S} in the context
of pulsar glitches. The following discussion provides additional, more technical,
details on these results.

Within the~\cite{2004MNRAS.347..575Y} approach all the initial
axisymmetric, non-corotating configurations of a corotating background
can be constructed as a linear combination of two independent classes
of initial data, see Section~\ref{sec:Noncorback}.  In the first
class, only the neutrons move relative to the corotating background,
i.e. $\left( \delta \Omega_\n , \delta \Omega_\p \right) = \left( 1 ,
0 \right)$, while in the second class only the proton velocity is
different from the corotating background, $\left( \delta \Omega_\n ,
\delta \Omega_\p \right) = \left( 0 , 1 \right)$. We will refer to these
two configurations as initial data N (ID-N) and P (ID-P), respectively.
The initial mass density $\delta \rho_x$, chemical potential $\delta
\tilde \mu_\x$ and gravitational potential $\delta \Phi$ can be
directly determined from equations~(\ref{eq:muxNC})--(\ref{eq:PoisNC})
for the two sets of initial data ID-N and ID-P. Meanwhile, for the velocity
field perturbation we consider
\begin{equation}
\qquad \delta \mtb{v}_\x =  \delta \Omega_\x  \, \left(  \mtb{\Omega}_\crt \times
\mtb{r} \right) \, .  \label{eq:vphi}
\end{equation}
These solutions can be rescaled to any required glitch size if we note
that the crust spin-up can be associated with the proton velocity lag
$\delta \Omega_\p$. In fact, we expect an efficient coupling between
the crust and the outer core protons due to the magnetic field, and we
can then assume that the charged particles corotate. The rotational lag between superfluid
neutrons and the
protons can be estimated by considering angular momentum conservation:
\begin{equation}
\delta J = I_\n \delta \Omega_{\n} + I_\p \delta \Omega_{\p} + \left( \delta
I_\n + \delta I_\n \right) \Omega_\crt = 0 \, . \label{eq:arel}
\end{equation}
Here $J$ is the total angular momentum, $I_\x$ is the moment of
inertia of each fluid constituent, and its perturbation $\delta I_\x$  is
defined by
\begin{equation}
\delta I_\x = \int_{0}^{\mtb{r}} \delta \rho_\x \, (r' \sin \theta )^2
d\mtb{r'} \, .
\end{equation}
The initial relative velocity lag that describes a glitch is then given by
\begin{eqnarray}
\delta \Omega_\p & = & \left. \frac{ \Omega_{\p} - \Omega_{\crt} }{ \, \, \Omega_\crt }\right|_{obs} \, , \label{eq:arel1}\\
\delta \Omega_{\n} & = & - \frac{1 }{I_\n }  \left(  I_\p\delta \Omega_{\p}
+ \delta I  \Omega_\crt \right) \,  , \label{eq:arel2}
\end{eqnarray}
where $\delta I = \delta I_\p + \delta I_\n$.

\begin{figure}
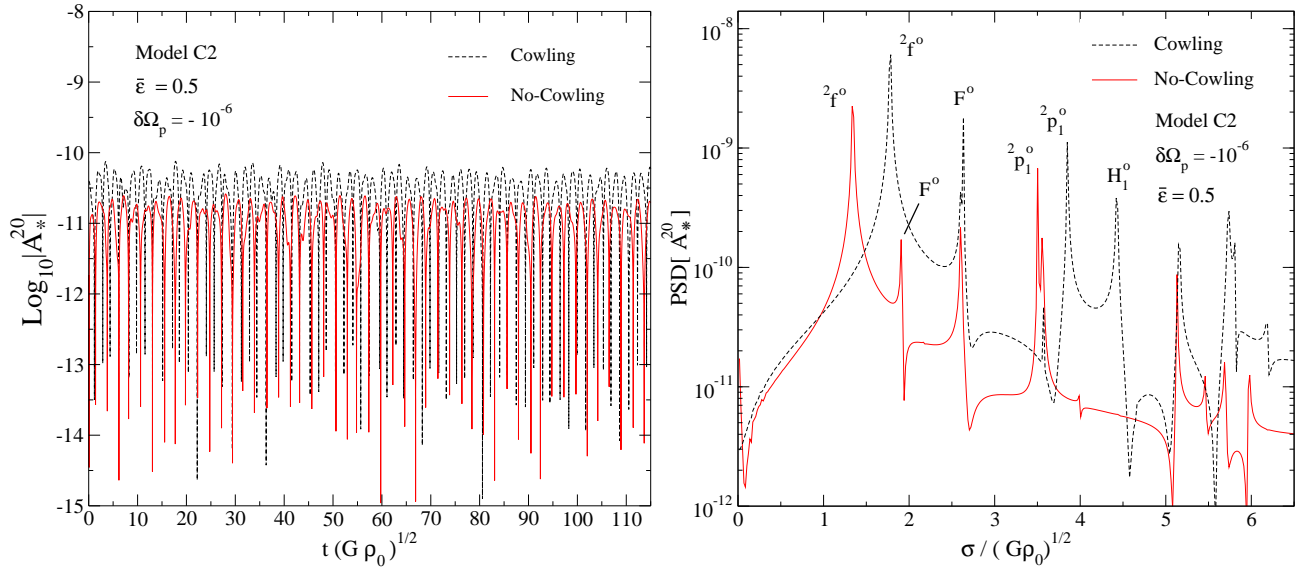

\begin{center}
\includegraphics[height=75mm]{fig12a.eps}
\includegraphics[height=74.2mm]{fig12b.eps}
\caption{In this figure we illustrate the effect of the Cowling
  approximation on the gravitational signal. The background star is
  the C2 model with $\bar \veps = 0.5$, and the initial condition is
  that used for the glitch model, i.e.~$\delta \Omega _\p = - 10^{-6}$
  and $\delta \Omega_\n = 7.74\times 10^{-8}$. The waveforms are shown
  in the left panel and the PSD in the right panel. The signals
  determined in Cowling approximation and with gravitational potential
  perturbation are shown with dashed and solid lines respectively. In
  the right panel we  note the effect of the Cowling
  approximation on the acoustic modes.  For the initial data
  considered in this work, the Cowling
  approximation model generates a larger gravitational-wave amplitude than
  in the case when the gravitational potential perturbation is accounted for.
  \label{fig:A20Cow}}
\end{center}
\end{figure}

For each background model, we can evolve  the two independent
initial data sets ID-N and ID-P. If we
consider a generic perturbation $\delta f$ for an arbitrary initial
configuration, we can determine the evolution from the following
linear combination:
\begin{equation}
\delta f = \delta f _{\rm{N}} \delta \Omega_\n + \delta f _{\rm{P}}
\Omega_\p \, ,
\end{equation}
where $\delta f _{\rm{N}}$ and $\delta f _{\rm{P}}$ are the
perturbation variables related to  ID-N and ID-P,
respectively.

We have evolved the ID-N and ID-P configurations for the C2 model with
$\bar \veps = 0.5$. In Fig.~\ref{fig:A20}, we show a part of the time
evolution of the quantity~$A^{20} = A^{20}_\n + A^{20}_\p$
determined from the stress-formula~(\ref{eq:A20-stres}). Actually, we
show the dimensionless quantity~$A^{20}_{\ast} = A^{20} / \left( G
\rho_{0}^{2} R_{eq}^{2} \right)$ that is directly determined by the
numerical code. For different values of the stellar parameters, the
gravitational-wave amplitude can be calculated from
equations~(\ref{eq:st-unit}) and~(\ref{eq:stc-unit}).  The initial
data ID-N generates a gravitational signal that is about an order of
magnitude larger than the ID-P initial data.  
We have studied
  stellar models with different proton fraction and noticed that the
  amplitude difference between the ID-N and ID-P initial data scales
  with the proton fraction of the background model. This is expected
  as the dynamics of the mass constituents generates the
  gravitational-wave signal.

For the glitch initial data, we study the effect of the Cowling
approximation on the gravitational-wave signal. To do this, we
consider two simulations for the same model C2 and entrainment
parameter $\bar \veps = 0.5$. The only difference is that, in one case
we neglect the perturbation of the gravitational potential~$\delta
\Phi$.  In Fig.~\ref{fig:A20Cow}, we show the time evolution of the
quantity $A^{20}_{\ast}$ and the related Power Spectrum Density (PSD),
which is defined as $\rm{PSD}(A^{20}_\ast) = | \hat{A}^{20}_\ast |$.
In the Cowling approximation, we extract the gravitational signal with
the momentum formula~(\ref{eq:A20-mom}), as the stress
formula~(\ref{eq:A20-stres}) is not well defined when $\delta \Phi =
0$. From the results in the left panel of Fig.~\ref{fig:A20Cow}, we
note that the Cowling approximation generates a signal that is about
five times larger than the result when $\delta \Phi$ is
included. Furthermore, as expected, the Cowling approximation
introduces a deviation in the mode frequencies.  This difference is
evident in the right panel of Fig.~\ref{fig:A20Cow}, where the error
is about $38\%$ for the fundamental quasi-radial mode
($\rm{F}^{\ord}$), $32\%$ for the axisymmetric $l=2$ f-mode
(${}^2\rm{f}^{\ord}$), and $10\%$ for the first pressure mode
(${}^2\rm{p}_{1}^{\ord}$). These results agree well with the results
of similar comparisons~\citep{1997MNRAS.289..117Y,
  2001MNRAS.322..389Y}. Regarding the amplitude of the
gravitational-wave signal, we find that the relative oscillation
amplitude between the Cowling approximation and the full problem
depends on the initial data. Hence, this result is not generic.

\section{Conclusions and Discussion\label{conclusions}} \label{sec:concl}

We have studied the dynamics of superfluid rotating neutron stars,
focussing on the nature of the oscillation spectrum, the effects of
the mutual friction force on the oscillations and the hydrodynamic
spin-up phase of pulsar ``glitches''. Adopting the Newtonian two-fluid
model, we evolved in time the perturbed dynamical equations on
axisymmetric equilibrium configurations. This approach allows us to
derive the spectrum of axisymmetric and non-axisymmetric oscillation
modes of stellar models that rotate up to the mass shedding limit. In
this work, we have improved on previous studies by including the
gravitational perturbation and the mutual friction force. The spectrum
is then determined with a better accuracy, as we no longer use the
Cowling approximation~\citep{2009MNRAS.396..951P}. From the
computational point of view, we have to solve the perturbed Poisson
equation together with the linearised momentum and mass conservation
equations. We have numerically evolved the hyperbolic equations with a
Mac-Cormack algorithm, while the elliptic equation for the
gravitational potential is solved at each time step with a spectral
method.

In our current model the rotating background models are pure fluid,
i.e. without an elastic crust region, neutrons and protons corotate
and are in $\beta$-equilibrium. In superfluid stars, the co- and
counter-phase motion of the two fluid constituents can be coupled by
composition gradients and this influences the dynamics. In order to
consider this effect we have studied two simple polytropic equations
of state that generate distinct sequences of stratified and
non-stratified rotating stars. These background models are simplistic,
and we must improve on this aspect if we want to decode the complexity
of astrophysical observations. Certainly, we must add an elastic crust
to the model and relax the co-rotation assumption between the two
fluids. If we want to use more realistic equations of state we also
need to translate the model to General Relativity. We are currently
working on all these issues.

In neutron stars, the mutual friction force may have both dissipative
and non-dissipative effects. The dissipative part of the force, which
is dominant in the weak drag regime, mainly damps an oscillation
mode. Meanwhile, the non-dissipative term dominates in the strong drag
regime, essentially modifying the oscillation spectrum.  We have
studied the two drag regimes and showed that our numerical code
effectively reproduces the mutual friction damping of the two-fluid
relative motion. For non-stratified stars, the co- and counter-moving
degrees of freedom are uncoupled and only the superfluid modes are
damped. When the stellar model is stratified, the damping affects also
the ordinary modes. The accuracy of our numerical code has also been
tested in~\cite{2010MNRAS.tmp..554S}, where determined the glitch
spin-up time and compared it to a simple analytic formula. However, we
are not yet able to extract (with useful precision) the mutual
friction damping time of individual oscillation modes from our
numerical evolutions. More work is needed to establish to what extent
one should expect to do this within our computational framework. For
the strong drag regime, we have studied the effect of the mutual
friction and composition variation on the rotational splitting of the
superfluid $l=m=2$ f-mode and on the frequencies of the $l=m=2$
superfluid r-mode. The main effect is a change of propagation
direction of the modes with respect to the background rotation. A mode
that is pro-grade (retro-grade) in the weak drag regime may become
retro-grade (pro-grade) in the strong drag regime. We have determined
the numerical frequencies of the f- and r-modes for the rotating
sequence of non-stratified stellar models and provided simple
empirical expressions based on the numerical data. For constant mutual
friction parameters, the non-axisymmetric splitting of the superfluid
f-mode and the r-mode frequencies depends on the inverse of the proton
fraction.

Finally, we provided relevant technical details for the hydrodynamical
models for pulsar glitches discussed
by~\cite{2010MNRAS.tmp..554S}. The initial conditions for the glitch
evolutions describe two fluids that rotate with a small velocity
lag. These configurations were been determined using a perturbative
approach first introduced by~\cite{2004MNRAS.347..575Y}. We extended
this method to implement different EoS and consider non-corotating
initial configurations that conserve the mass of each fluid
constituent. Moreover, we derived the detailed quadrupole
gravitational extraction formulae for $l=2$ oscillation modes of a
superfluid star. We determined the perturbative expressions for the
momentum and stress formulae that can be used to improve the numerical
extraction of the gravitational-wave signal (reducing the order of the
time derivative of the standard quadrupole formula). We determined the
gravitational-wave strain for the two independent initial glitch
configurations that are obtained with the~\cite{2004MNRAS.347..575Y}
approach. For a given background rotation, these results can be used
to estimate the gravitational signal for any glitch size. Furthermore,
we have showed the effect of the Cowling approximation on the glitch
gravitational-wave strain and the oscillation spectrum.

With the progress described in this paper, our programme of studying
superfluid neutron star dynamics by time-evolutions of the linearised
equations has reached the point where we need to add key physics to
the model.  The natural step would be to account for the elastic
neutron star crust with the expected interpenetrating neutron
superfluid. This requires us to change the computational framework
somewhat, as it is natural to discuss the elasticity in term of
Lagrangian perturbation theory. Moreover, we need to address various
issues associated with vortex pinning by the crust nuclei. This
problem requires additional force contributions at the level of
individual vortices, and we need to develop a suitable smooth-averaged
hydrodynamics description if we want to make progress. We are
currently working on both these issues.  It would also be relevant to
extend our models to general relativity.  This is essential if we want
to be able to use realistic supranuclear equations of state.  As long
as we make use of the relativistic analogue of the Cowling
approximation this generalisation should be straightforward, but if we
want to account for the dynamics of spacetime the problem becomes much
more involved.  If we want to consider realistically ``layered''
neutron stars we also need to improve our understanding of the
different phase-transitions, e.g. in the vicinity of the critical
density/temperature for the onset of superfluidity, and how these
regions affect the large scale dynamics. We face a number of
challenging questions, but there is no reason why we should not be
able to resolve the relevant issues and progress towards the
construction of realistic dynamical neutron star models.

\section*{Acknowledgements}
This work was supported by STFC through grant number PP/E001025/1.

\appendix

\section{GW extraction} \label{sec:gwextr}

In this Appendix we determine the momentum formula~(\ref{eq:A22-mom})
and the stress formula~(\ref{eq:A22-stress}) for the $(l,m)=(2,2)$
gravitational signal.  For the axisymmetric $(l,m)=(2,0)$ case, we
have used the perturbative version of the momentum and stress formulae
used by~\cite{1990ApJ...351..588F}.

The aim is to reduce the order of the time derivatives in the
quadrupole gravitational-wave formula. To do this we consider the
quantity~(\ref{eq:hp3a}):
\begin{equation}
A^{22} \equiv \frac{d^2}{dt^2} \int d\mtb{r} \, \delta \rho \, r^2
Y_{22}^{\ast} \, .  \label{eq:A22def}
\end{equation}
With the use of the mass conservation equations of each fluid
component:
\begin{equation}
\partial_t \rho_{\x} + \nabla_{i} \left( \rho_{\x} v_{\x}^{i} \right)
= 0 \, , \label{eq:MconA}
\end{equation}
we can determine the momentum-formula where only a first order time
derivative appears. When we perturb equation~(\ref{eq:MconA}) and introduce it
in~(\ref{eq:A22def}) we obtain:
\begin{eqnarray}
A^{22} & = &\frac{d^2 }{dt^2} \int d\mtb{r} \, \delta \rho \, r^2 \,
Y_{22}^{\ast} = \sum_{\x} \frac{d}{dt} \int d\mtb{r} \, \partial_t
\left( \delta \rho_{\x} \right) \, r^2 \, Y_{22}^{\ast} = - \sum_{\x}
\frac{d}{dt} \int d\mtb{r} \, \nabla_{i} \delta \left( \rho_{\x}
v_{\x}^{i} \right) r^2 \, Y_{22}^{\ast} \nn \\ {} & = & \sum_{\x}
\frac{d}{dt} \int d\mtb{r} \, \, \delta \left( \rho_\x v^{k}_\x
\right) \nabla_{k} \left( r^2 \, Y_{22}^{\ast} \right) \, .
\label{eq:mom-fr}
\end{eqnarray}
where in the last step we have used the Gauss Theorem. After some
calculation, equation~(\ref{eq:mom-fr}) leads to the
expression~(\ref{eq:A22-mom}).

With a similar method, we can determine the stress-formula and
eliminate the time derivatives from the quadrupole formula. In this
case, we must use the momentum conservation equation that for a
superfluid component is given by
\begin{equation}
\partial_t \left( \rho_{\x} p^{\x}_{i} \right)
+  \nabla_{k} \left( \rho_{\x} v_\x ^{i} p^{\x}_{i}  \right)
+ \rho_{\x} \nabla_{i} \tilde{\mu}_{\x} + \rho_{\x} \nabla_{i} \Phi
+ \rho_\x \veps_\x w_{k}^{\y\x} \nabla_{i} v_\x^{k} = 0
\, ,  \label{eq:EulerA}
\end{equation}
where the momentum of the fluid component is defined as follows:
\begin{equation}
p_{i}^{\x} = v_{i}^{\x} + \varepsilon_{\x} w_{i}^{\y\x} \, .
\end{equation}
For a two-fluid model with neutron and proton as components, the total
momentum equation is then given by the following expression:
\begin{equation}
\partial_t \left( \rho_{\n} v^{\n}_{i} + \rho_{\p} v^{\p}_{i} \right) =
-   \nabla_{k} \left( \rho_{\n} v_\n ^{k} p^{\n}_{i} + \rho_{\p} v_\p ^{k} p^{\p}_{i}  \right)
-  \nabla_{i} \Psi - \frac{1}{4\pi G} \nabla^{k} \left( \nabla_{k} \Phi \nabla_{i} \Phi
- \frac{\delta_{ik}}{2} \nabla_{j} \Phi \nabla^{j} \Phi \right)
\, ,  \label{eq:EulerB} \\
\end{equation}
where we have used the definition of the the generalized pressure~\citep{2004PhRvD..69d3001P}:
\begin{equation}
\nabla  \Psi = \rho_{\n} \nabla \tilde{\mu}_\n + \rho_{\p} \nabla \tilde{\mu}_\p - \frac{1}{2} \rho_\x \veps_\x  \nabla \left( w_{\p\n}^2\right)  \, ,
\end{equation}
and we have re-written the gravitational potential term by using the
Poisson equation:
\begin{equation}
\rho \nabla_{i} \Phi = \frac{1}{4\pi G} \nabla^{k} \left( \nabla_{k} \Phi \nabla_{i} \Phi
- \frac{\delta_{ik}}{2} \nabla_{j} \Phi \nabla^{j} \Phi \right) \, .
\end{equation}
Perturbing equation~(\ref{eq:EulerB}) and considering a corotating
equilibrium configuration, i.e. $w_{\n \p} = 0$, we obtain:
\begin{equation}
\frac{\partial}{\partial t} \, \delta \left( \rho_{\n} v^{\n}_{i} + \rho_{\p} v^{\p}_{i} \right) =
-   \nabla_{k} \delta \left( \rho_{\n} v_\n ^{k} p^{\n}_{i} + \rho_{\p} v_\p ^{k} p^{\p}_{i}  \right)
-  \nabla_{i} \delta P
- \frac{1}{4\pi G} \nabla^{k} \delta \left( \nabla_{k} \Phi \nabla_{i} \Phi - \frac{\delta_{ik}}{2}
\nabla_{j} \Phi \nabla^{j} \Phi \right)
\, ,  \label{eq:EulerC} \\
\end{equation}
where now for corotating background the pressure perturbation is given
by
\begin{equation}
\nabla \delta P = \delta \left( \rho_\n \nabla \tilde \mu _\n + \rho_\p \nabla \tilde \mu _\p \right) \, .
\end{equation}
We can now introduce equation~(\ref{eq:EulerC}) into
equation~(\ref{eq:mom-fr}) and use the Gauss theorem. We obtain:
\begin{equation}
\frac{d^2 }{dt^2} \int d\mtb{r} \, \delta \rho \, r^2 \, Y_{22}^{\ast}
=  \int d\mtb{r} \, \delta \left( \rho_{\n} v_\n ^{k} p^{i}_{\n} + \rho_{\p} v_\p ^{k} p^{i}_{\p}
+ \frac{1}{4\pi}   \nabla^{k} \Phi \nabla^{i} \Phi \right)  \nabla_{i} \nabla_{k}  \left( r^2 \,
 Y_{22}^{\ast} \right)
\label{eq:trB}
\end{equation}
where both the pressure and the last term of
equation~(\ref{eq:EulerC}) vanish, as $\nabla^2 \left( r^2 \,
Y_{22}^{\ast} \right) = 0$. After some further calculation,
we can derive equation~(\ref{eq:A22-stress}) from~(\ref{eq:trB}).

\nocite*
\bibliographystyle{mn2e}

\label{lastpage}

\end{document}